\documentclass[twocolumn,showpacs]{revtex4}

\usepackage{dcolumn}
\usepackage{bm}
\usepackage{graphicx}
\usepackage{verbatim}
\usepackage{epsf}

\newcommand{\ve}[1]{\mbox{\boldmath$#1$}}

\newcommand{\pt}{\ensuremath{p_{\rm t}}}

\begin{document}

\title{Non-equilibrium hadronization and\\ constituent quark number scaling}

\author{\sc Sven Zschocke$^{\bf 1}$, Szabolcs Horv\'at$^{\bf 2}$,  
Igor N. Mishustin$^{\bf 3,4}$, L\'aszl\'o P. Csernai$^{\bf 2,3,5}$}
\affiliation{$^{\bf 1}$ TU Dresden, Institut f\"ur Planetare Geod\"asie, 
Lohrmann-Observatorium, D-01062 Dresden, Germany
\\
$^{\bf 2}$ Department of Physics and Technology, University of Bergen, Allegaten 55, N-5007 Bergen, Norway
\\
$^{\bf 3}$ Frankfurt Institute for Advanced Studies (FIAS),
 D-60438 Frankfurt am Main, Germany
\\
$^{\bf 4}$ Russian Research Center, Kurchatov Institute,
123182 Moscow, Russia
\\
$^{\bf 5}$ MTA--KFKI Research Inst.\ for Particle Physics and Nuclear 
Physics, 1525 Budapest, Hungary} 

\begin{abstract}
The constituent quark number scaling of elliptic flow is studied in a
non-equilibrium hadronization and freeze-out model with rapid dynamical
transition from ideal, deconfined and chirally symmetric Quark Gluon Plasma,
to final non-interacting hadrons. In this transition a Bag model of
constituent quarks is considered, where the quarks gain constituent quark
mass while the background Bag-field breaks up and vanishes. The
constituent quarks then recombine into simplified hadron states, while
chemical, thermal and flow equilibrium break down one after the other.
In this scenario the resulting temperatures and flow velocities of 
baryons and mesons are different.
Using a simplified few source model of the elliptic flow,
we are able to reproduce the constituent quark number scaling,
with assumptions on the details of the non-equilibrium processes.
\end{abstract}


\maketitle


\section{Introduction}\label{intro}

It has been observed that the momentum distribution of particles
created in heavy ion collisions is not azimuthally
symmetric in the plane perpendicular to the beam direction.
This dominant asymmetry, the elliptic flow, is the result of several
factors, the main one being the anisotropy of the
initial configuration after the collision due to the
non-zero impact parameter.
The elliptic flow is typically characterized using the second
coefficient of the Fourier expansion of the momentum
distribution, $v_2$. It was found in experiments
\cite{CQNS} that the $v_2$
parameter as a function of the transverse momentum, $\pt$,
scales with the number of constituent quarks, $n_{cq}$, in a hadron. That is, 
if the $v_2(\pt)$ curves are re-scaled according to the constituent
quark number of the considered hadron species, 
and $v_2/n_{cq}$ 
is plotted as function of $\pt/ n_{cq}$ for each type of hadron with mass $m_h$, 
the curves will coincide. Later results showed that the
scaling is more precise if $v_2$ is plotted as a function of 
the transverse kinetic energy,
$E_{\rm t} = \sqrt{m_h^2 + \pt^2} - m_h$, instead of the transverse momentum \pt.

This experimentally found scaling law is remarkably simple, and indicates that the elliptic flow develops before the 
quarks recombine into hadrons. Therefore, understanding the factors influencing the measurable $v_2$ can provide 
information about the state of the Quak Gluon Plasma (QGP).

The fluid dynamical (FD) model describes the dynamical development of the 
QGP from the (already thermalized) initial state
until the break-down of the equilibrium, where first the chemical
equilibrium among quarks and anti-quarks ceases. Initially the
plasma has only two flavors, u and d, and thus the FD model assumes
two flavors. However, the flavor equilibration in the plasma is
a rapid process with a timescale of the order of $1\,{\rm fm}/c$, so by the
end of the FD development a flavor equilibrium among three flavors,
u, d, and s, is reached.
       Our usual FD evolution does not take into account this chemical
change of flavors with an additional rate equation, instead this
process is taken into account at the final stage of the FD development,
when the subsequent EoS is already assumed to have three equilibrated
flavors. Energy and momentum conservation and the requirement of
non-decreasing entropy is enforced in the transition, from the ideal QGP 
state to the state where quark and anti-quark numbers
are frozen out.
       As a consequence, the mass change of the quarks starts in the
initial QGP with two flavors, and we use this approximation to
estimate in Section II.B the final boundary where the initial
FD stage of the evolution ends, with light quark masses.

We consider a model of hadronization and investigate the constituent 
quark number scaling (QNS) of the $v_2$ parameter.
In this model, a gas of quarks and anti-quarks expands in a background field
represented by the Bag constant $B$, which depends on density and temperature 
of the expanding fireball. Initially this $B$-field includes the energy of 
the deconfined perturbative vacuum and of the gluon fields. 
As the system expands the deconfinement starts
and the average $B$ decreases. Furthermore, as the chiral symmetry breaking
starts, the quarks gain mass. The quark mass is calculated as a 
function of the temperature and density of the matter.
The quark gas expands rapidly while the quark mass increases. 
This process can be considered as a simple representation of the breaking chiral symmetry
and deconfinement in a dynamical transition crossing the Quarkyonic phase \cite{McLerran_Pisarski}.

The point during the expansion when the quarks recombine into
hadrons is determined from the condition that the average hadron energy
is equivalent to $\left(1.0 - 1.1\right){\rm GeV}$, as found from the systematics of 
experimental data \cite{Cleymans}. 
At recombination, the thermal and flow equilibrium between particles is broken.

Finally, the $v_2$ parameter is determined using simple two- and three-source models of the elliptic flow and the particle
distributions obtained from the hadronization model.

The paper is organized as follows:
In Section \ref{A} the density and temperature dependence of constituent quark mass is considered. The energy, entropy,
momentum and chemical potential of each individual source and of the total source is given in Section \ref{B}.
In Section \ref{C} we describe the applied non-equilibrium model of rapid hadronization. 
Details of calculation of elliptic flow parameter $v_2$ are presented in Section \ref{D}.
A summary is given in Section \ref{E}.

\section{Constituent quark mass}\label{A}

To present our arguments we consider the Nambu-Jona-Lasinio model (NJL) which is motivated by 
Quantum Chromodynamics (QCD) and is basically 
a purely quark-quark-interaction theory; a comprehensive overview about NJL has been presented in \cite{Klevansky}. 
The extended NJL-model contains three types of quark-quark-interaction and is given by the Lagrangian
\begin{eqnarray}
&& {\cal L}_{\rm NJL} = \sum\limits_{f=u,d,s} \overline{q}_f \left(i\,\gamma^{\mu} \partial_{\mu} - m_f \right) q_f
\nonumber\\
\nonumber\\
&& + \frac{G_S}{2} \sum\limits_{f=u,d,s}\sum\limits_{a=0}^{8} \left[\left(\overline{q}_f \lambda^a q_f\right)^2
+ \left(\overline{q}_f \, i \,\gamma_5\,\lambda^a q_f\right)^2 \right]
\nonumber\\
\nonumber\\
&&-\frac{G_V}{2}\sum\limits_{f=u,d,s}\sum\limits_{a=1}^{8}\left[\left(\overline{q}_f\,\gamma_{\mu}\,\lambda^a q_f\right)^2
+ \left(\overline{q}_f \,\gamma_5\,\gamma_{\mu}\,\lambda^a q_f\right)^2 \right]
\nonumber\\
\nonumber\\
&& + \frac{G_D}{2} \left[ {\rm det} 
\left( \overline{q}_f ( 1 + \gamma_5) q_f \right) + {\rm det} \left( \overline{q}_f ( 1 - \gamma_5) q_f \right) \right],
\label{Eq_12}
\end{eqnarray}

\noindent
where $\overline{q}_f = q_f^{\dagger}\,\gamma_0$, $\gamma^{\mu}$ are the Dirac matrices, 
$m_f$ is the current quark mass of flavor $f$, 
$G_S$ is the coupling constant of the scalar-current interaction, 
$G_V$ is the coupling constant of the vector-current interaction terms and $G_D$ is the coupling constant of the 
determinantal flavor-mixing term (determinant in flavor space). The Gell-Mann matrices are $\lambda^a$ where 
$a = 0,1,2,\dots,8$ with $\lambda^0 = \sqrt{2/3}\,I$, where $I$ is the unit matrix. This Lagrangian is used to derive the 
relation between constituent quark masses and chiral condensates, both for vacuum and for finite densities and temperatures.

\subsection{Constituent quark mass in vacuum}

First, we consider the constituent quark mass in vacuum. For three flavors, $SU_f (3)$, the constituent quark masses 
in vacuum $M_f^0 = M_f \left(n_B=0,T=0\right)$ where $f=u,d,s$, are related to the ciral condensates in vacuum 
as follows \cite{Vogel_Weise,MishustinCQ,LiShakin} 
\begin{eqnarray}
M_u^0 &=& m_u - 2\;G_S\;\langle \overline{u} u \rangle_{0} - G_D\;  \langle \overline{d} d \rangle_{0} \;\langle \overline{s} s \rangle_{0} \,,
\label{Eq_23}
\\
M_d^0  &=& m_d - 2\;G_S\;\langle \overline{d} d \rangle_{0} - G_D\;  \langle \overline{u} u \rangle_{0} \;\langle \overline{s} s \rangle_{0} \,,
\label{Eq_24}
\\
M_s^0 &=& m_s - 2\;G_S\;\langle \overline{s} s \rangle_{0} - G_D\;  \langle \overline{u} u \rangle_{0} \;\langle \overline{d} d \rangle_{0} \,.
\label{Eq_25}
\end{eqnarray}

\noindent
Here $\overline{u} = u^{\dagger}\,\gamma_0$ and so on, 
$m_u = 5\,{\rm MeV}, m_d = 9\,{\rm MeV}$ and $m_s = 130\,{\rm MeV}$ are the current quark masses of the u-quark, 
d-quark and s-quark, respectively. Typical values of the chiral condensates in vacuum 
are given by, e.g. \cite{Cohen_Furnstahl1,Cohen_Furnstahl2} and references therein:
\begin{eqnarray}
\langle \overline{q} q \rangle_{0} &=&\langle \overline{u} u \rangle_{0} = \langle \overline{d} d \rangle_{0} = - (0.225\;{\rm GeV})^3\,,
\label{Eq_30}
\\
\langle \overline{s} s \rangle_{0} &=& 0.7\;\langle \overline{u} u \rangle_{0}\,.
\label{Eq_35}
\end{eqnarray}

\noindent
Typical values for the coupling constants are $G_S = (15 - 20) \;{\rm GeV}^{-2}$, $G_V \simeq 0.5\,G_S$ and 
$G_D = - (160 - 240)\;{\rm GeV}^{-5}$ \cite{Klevansky,MishustinCQ,LiShakin}. 
We note that in relations (\ref{Eq_23}) - (\ref{Eq_25}) the magnitude of terms proportional to $G_D$ 
are small compared to the terms proportional to $G_S$. 

\subsection{Constituent quark mass in a hot and dense medium}

Let us now consider the case of quarks in a hot and dense medium,
at early times of FD evolution. At this stage we assume to have two
flavors. Then, the relations (\ref{Eq_23}) - (\ref{Eq_25}) for two flavors $SU_f (2)$ are reduced to 
\begin{eqnarray}
M_f &=& m_f - 2\;G_S\;\langle \overline{q} q \rangle_{n_B,T}\,.
\label{Eq_50}
\end{eqnarray}

\noindent 
The suffix $n_B,T$ at chiral condensate denotes the Gibbs average over hadron states and meson states 
\cite{Bochkarev,HatsudaKoikeLee,zschocke1,zschocke3} and references therein. Note that for isospin-symmetric matter there 
is no difference in the density and temperature dependence of u-quark and d-quark condensate, that means 
$\langle\overline{q}q \rangle_{n_B,T}\equiv\langle\overline{u}u \rangle_{n_B,T}=\langle \overline{d}d \rangle_{n_B,T}$. 
In the limit of high densities and temperatures the constituent quark mass, $M_f$, approaches the current quark mass, $m_f$. 
The well-known model-independent linear density dependence of the chiral condensate has widely been applied 
in many investigations, e.g. \cite{Drukarev_Levin,LiShakin,Cohen_Furnstahl1,zschocke1}. The temperature dependence of the 
chiral condensate has been determined up to order ${\cal O} (T^8)$ in \cite{Gerber_Leutwyler}.
In order to determine the density {\it and} temperature dependence of the chiral condensate we follow 
the arguments of Refs.~\cite{LiShakin,zschocke1,zschocke2}, where the first leading terms in the low-density 
low-temperature expansion have been obtained:
\begin{eqnarray}
&& \langle \overline{q} q \rangle_{n_B,T} = \langle \overline{q} q \rangle_{0} 
\nonumber\\
&& \times \left(1 - \frac{3\,\sigma_q}{f_{\pi}^2\;m_{\pi}^2}\,n_B
- \frac{T^2}{8\,f_{\pi}^2} - \frac{T^4}{384\,f_{\pi}^4} - \frac{T^6}{288\,f_{\pi}^6}\,\ln \frac{\Lambda_q}{T} \right).
\nonumber\\
\label{Eq_60}
\end{eqnarray}

\noindent
The temperature and density dependence of chiral condensate has been plotted in Ref.~\cite{zschocke2}.
The baryonic density in terms of quark degrees of freedom is given by
\begin{eqnarray}
n_B &=& \frac{1}{3} \sum\limits_{f=u,d} \left( n_f - \overline{n}_f \right)\,,
\label{baryonic_density_1}
\end{eqnarray}

\noindent
where $n_f$ ($\overline{n}_f$) is the quark (anti-quark) density. The baryonic density in a given volume $V$ is related to 
the conserved baryon number $N_B$ by $n_B = \frac{\displaystyle N_B}{\displaystyle V}$; note the relation 
$\hbar\,c = 197.3\,{\rm MeV}\,{\rm fm}$. For the logarithmic scale we take $\Lambda_q \simeq 300\;{\rm MeV}$, for details 
see \cite{Gerber_Leutwyler}. The pion mass in vacuum is $m_{\pi} = 138\;{\rm MeV}$ and the pion decay constant in vacuum 
is $f_{\pi} = 93\;{\rm MeV}$. The numerical value of the quark-sigma-term is $\sigma_q = 15\;{\rm MeV}$ (see e.g.\ Ref.~\cite{LiShakin}), which is  three times smaller than the nucleon-sigma-term $\sigma_N = 45\;{\rm MeV}$.

\begin{figure}[!h]
\begin{center}
\includegraphics[scale=0.3]{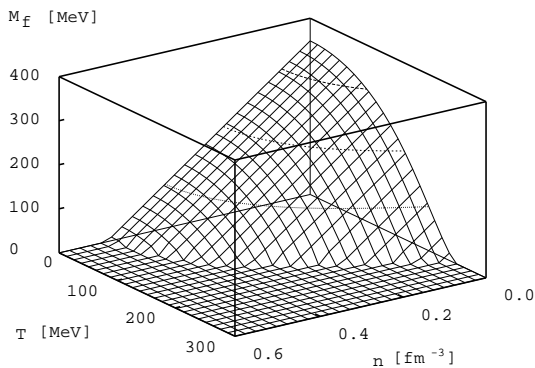}
\caption{Temperature and density dependence of the constituent quark mass $M_f$ according to Eq.~(\ref{Eq_70}).}
\label{condensate}
\end{center}
\end{figure}

\noindent
By combining Eq.~(\ref{Eq_50}) with Eq.~(\ref{Eq_60}), we obtain the expression for the in-medium mass of 
constituent-quarks:
\begin{eqnarray}
&& M_f = m_f - 2\,G_s \langle \overline{q} q \rangle_0  
\nonumber\\
&&\times \left(1 - \frac{3\,\sigma_q}{f_{\pi}^2\,m_{\pi}^2}\,n_B
- \frac{T^2}{8\,f_{\pi}^2} - \frac{T^4}{384\,f_{\pi}^4} - 
\frac{T^6}{288\,f_{\pi}^6}\,\ln \frac{\Lambda_q}{T} \right).
\nonumber\\
\label{Eq_70}
\end{eqnarray}

\noindent
The temperature and density dependence of the constituent quark mass is plotted in FIG.~\ref{condensate}.
According to Eq.~(\ref{Eq_70}), for sufficiently high temperatures and densities the constituent quark mass will coincide 
with the current-quark-mass $m_f$. Of course, the applicability of Eq.~(\ref{Eq_70}) is restricted to low densities and 
temperatures. In any case, the most upper limit for density and temperature is given by the condition that constituent 
quark mass has to be positive.

\section{Expanding quark--anti-quark bag}\label{B}

The Bag consists of quarks, anti-quarks, and the gluon fields. 
Non-perturbative effects in the QCD vacuum and the energy density of 
gluon fields are described by the Bag constant, $B$.  
In our model, we subdivide the bag into $N$ individual cells, 
each of which moves with an individual flow velocity $\ve{v}_i$.

\subsection{Total energy of the bag}

The total energy of the bag in the center-of-mass frame of the
colliding nuclei is given by the volume integral of the "00" component
of the energy momentum tensor of each cell $i$,
$T_i^{\mu\nu} = \left(e^i+P^i\right) u^{\mu}_i u^{\nu}_i - P^i\,g^{\mu\nu}$.
Considering that our EoS is given as a sum of the energy and pressure of the
ideal quark and anti-quark gas plus the uniform bag energy density $B$, that means 
$e^i = e^i_q + {\overline e}^i_q + B$ and $P^i = P^i_q + {\overline P}^i_q - B$, 
the total energy of all cells of the bag in the center-of-mass frame of the colliding nuclei is given by
\begin{eqnarray}
E_{\rm total} &=& \sum\limits_{i=1}^N\,V^i\,
\left(\gamma^i\right)^2 \left(e^i + P^i\right) 
- \sum\limits_{i=1}^N \, V^i \, P^i\,,
\label{total_energy}
\end{eqnarray}

\noindent
where the sum runs over the number of all cells of the bag. 
Here, we assume a uniform bag-field energy density $B$
over the whole volume, $V$; $B$ depends on 
density and temperature of the bag, thus
on the time of the evolution of the fireball.
We have used the notation $e^i$ for the invariant scalar, rest-frame 
energy density, and $P^i$ for the scalar pressure, $V^i$ 
is the volume of cell $i$ in the center-of-mass frame, 
and $\gamma_i = 1/\sqrt{1-{\bf v}_i^2}$ where ${\bf v}_i$ is 
the 3-velocity of cell $i$.  In order to determine the rest-frame
energy or proper energy of the cells, we assume a 
Boltzmann-J\"uttner distribution
\cite{Juttner} for the particles inside each individual cell:
\begin{eqnarray}
f_{\rm J}^i &=&
\frac{1}{\left(2\,\pi\,\hbar\right)^3} \;{\rm exp} 
\left(\frac{\mu^i - u_{\mu}^i\;p^{\mu}}{T_i}\right) \,,
\label{Juttner_A}
\end{eqnarray}

\noindent
and $\int d^3 x\;\int d^3 p \;f_i (\ve{x}, \ve{p}) = N_i$ is the normalization of the J\"uttner distribution, where $N_i$ 
is the number of particles of a given type inside cell $i$ under consideration. The proper energy density is a Lorentz 
invariant quantity by definition and, therefore, can be evaluated in any frame, e.g. in the local rest frame. Accordingly, 
we obtain the following expression for the energy density of quarks and anti-quarks for each cell of the bag:
\begin{eqnarray}
e^i &=& e_q^i + {\overline e}_q^i + B\,,
\label{Eq_75}
\\
e_q^i &=& \frac{1}{8\;\pi^2}\;\sum\limits_{f=u,d}\;
M_f^3 \;T_i\;{\rm exp} \left(\frac{\mu_q^i}{T_i}\right)
\nonumber\\
&& \times \left[K_1 \left(\frac{M_f}{T_i}\right) + 3 K_3 
\left(\frac{M_f }{T_i}\right)\right], 
\label{Eq_80} 
\\
{\overline e}_q^i &=& \frac{1}{8\;\pi^2}\;\sum\limits_{f=u,d}\,
M_f^3\,T_i\,\exp \left(\frac{{\overline \mu}_q^i}{T_i}\right)
\nonumber\\
&& \times \left[K_1 \left(\frac{M_f }{T_i}\right) + 3 K_3 
\left(\frac{M_f }{T_i}\right)\right].
\label{Eq_85}
\end{eqnarray}

\noindent
Here, the density and temperature dependent constituent quark mass $M_f$ is given in (\ref{Eq_70}), 
and $n_B^i$ and $T_i$ are the baryonic density and temperature, respectively, of the cell $i$.
The initial value of the Bag constant is abbreviated by $B_0$, and 
we use the numerical value
$B_0 = \left(198\;{\rm MeV}\right)^4 = 0.2\;{\rm GeV}\;{\rm fm}^{-3}$; 
see also considerations in Ref. \cite{Szhorvat}.

\subsection{Total entropy of the bag}

The total entropy in the center-of-mass Lorentz frame (CM) is given by
\begin{eqnarray}
S_{\rm total} &=& \sum\limits_{i=1}^N\;V^i \;\gamma^i\;s^i\,,
\label{total_entropy}
\end{eqnarray}

\noindent
where the sum runs over all cells of the bag. The invariant scalar entropy density, $s^i$, can be evaluated in any frame. 
The total entropy density consists of entropy density of the quarks and anti-quarks, and for the assumed J\"uttner 
distribution is given by:

\newcommand{\mf}{\ensuremath{M_f}}
\begin{eqnarray}
s^i &=& s^i_q + {\overline s}^i_q \,,
\label{entropy_10}
\\ 
s_q^i &=& \frac{1}{2\pi^2} \sum_{f=u,d} M_f^2 \exp\left(\frac{\mu_q^i}{T_i}\right) 
\nonumber\\
&&\hspace{-2.3cm} \times \left[M_f\, K_1\left(\frac{\mf}{T_i}\right) +
T_i \left(4 - \frac{\mu_q^i}{T_i}\right) 
K_2 \left(\frac{\mf}{T_i}\right) \right],
\label{entropy_11}
\\
{\overline s}_q^i &=&
\frac{1}{2\pi^2} \sum_{f=u,d}
M_f^2 \exp\left(\frac{{\overline \mu}_q^i}{T_i}\right) 
\nonumber\\
&&\hspace{-2.3cm} \times \left[M_f\, K_1\left(\frac{\mf}{T_i}\right) +
T_i \left(4 - \frac{{\overline \mu}_q^i}{T_i}\right) 
K_2 \left(\frac{\mf}{T_i}\right) \right].
\label{entropy_12}
\end{eqnarray}

\noindent
In what follows, we will take the adiabatic limit, which implies the 
total entropy to be a constant. We note that there is no Bag constant 
in the expression for the total entropy because the Bag-field is 
uniform, i.e. the entropy of Bag-field vanishes.

\subsection{Total momentum of the bag}

The total momentum of the bag in the center-of-mass Lorentz-frame is given by
\begin{eqnarray}
0 = \ve{p}_{\rm total} &=& 
\sum\limits_{i=1}^N\;V^i \;\left(\gamma^i\right)^2 \;
\left(e^i + P^i \right) \ve{v}_i\,,
\label{total_momentum}
\end{eqnarray}

\noindent
where the sum runs over all cells of the bag. 
The scalar pressure of each individual cell of the bag is given by
\begin{eqnarray}
P^i &=& P_q^i + {\overline P}_q^i - B \,,
\label{Eq_96}
\\
P_q^i &=& \frac{1}{2\,\pi^2}\,\sum\limits_{f=u,d}\;
M_f^2\,T_i^2\,
{\rm exp} \left( \frac{\mu_q^i}{ T_i}\right) K_2 
\left(\frac{M_f}{T_i}\right),
\nonumber\\
\label{Eq_101}
\\
{\overline P}_q^i &=& \frac{1}{2\,\pi^2}\;\sum\limits_{f=u,d}\,M_f^2\,T_i^2\,
{\rm exp} \left( \frac{{\overline \mu}_q^i}{ T_i}\right) 
K_2 \left(\frac{M_f }{T_i}\right).
\nonumber\\
\label{Eq_102}
\end{eqnarray}

\noindent
We note, that there is no Bag constant in the expression for the total momentum (\ref{total_momentum}) because 
it cancels out in the sum $e^i + P^i$ according to Eqs.~(\ref{Eq_75}) and (\ref{Eq_96}).

\subsection{Chemical potential of quarks and anti-quarks}

Assuming a J\"uttner distribution, the density $n_f^i = \frac{\displaystyle N_f^i}{\displaystyle V_i}$ of quarks having 
flavor $f$ and the density of anti-quarks $\overline{n}_f^i = \frac{\displaystyle \overline{N}_f^i}{\displaystyle V_i}$ 
having flavor $f$ in a given cell $i$ is determined by the following equations:
\begin{eqnarray}
n_f^i &=& \frac{1}{2\,\pi^2}\,M_f^2\,T_i\,{\rm exp} 
\left(\frac{\mu_q^i}{T_i}\right) K_2 \left(\frac{M_f }{T_i}\right),
\label{chemical_potential_5}
\\
\overline{n}_f^i &=& \frac{1}{2\,\pi^2}\,M_f^2\,
T_i\,{\rm exp} \left( \frac{{\overline \mu}_q^i}{T_i}\right) K_2 \left(\frac{M_f }{T_i}\right).
\label{chemical_potential_10}
\end{eqnarray}

\noindent
In mechanical (pressure), thermal and chemical equilibrium, the number of quarks depends only on baryonic chemical 
potential $\mu_B$ and temperature $T$. Especially, if the quarks and anti-quarks are in chemical equilibrium with each 
other, then $\mu_q = \mu_B / 3$ and ${\overline \mu}_q = -\mu_B/3$. 
In order to determine the baryonic chemical potential $\mu_B$ 
we have to use the total baryonic number density $n_B$ given in Eq.~(\ref{baryonic_density_1}). Then, according to the 
expressions (\ref{chemical_potential_5}) and (\ref{chemical_potential_10}), we obtain a definition of the chemical 
potential $\mu_B^i$ of cell $i$ as 
\begin{eqnarray}
n_B^i &=& \frac{1}{3\,\pi^2}\,T_i\,\sinh \left(\frac{\mu_B^i}{3\,T_i}\right)
\sum\limits_{f=u,d} M_f^2\,K_2 \left(\frac{M_f}{T_i}\right),\,
\label{baryonic_density_2}
\end{eqnarray}

\noindent
where we have used $\sinh x = \frac{1}{2} \left(e^x - e^{-x}\right)$. 
We assume that these thermodynamic relations hold in the ideal QGP until the final stage of the expansion starts. 
Then the net baryon number in each cell $N^i_B$ and the number of quarks and anti-quarks, $N_q^i = (n_u^i + n_d^i) V_i$ 
and ${\overline N}_q^i = ({\overline n}_u^i + {\overline n}_d^i) V_i$, respectively, are given. 
While $V_i$ and thus $n_f^i$ and ${\overline n}_f^i$ are changing during the expansion, 
the numbers of quarks and anti-quarks, $N_q^i$ and ${\overline N}_q^i$, are 
fixed at this point in each cell and do not change afterwards. This is the chemical freeze out in the model.

The volumes, baryonic and quark densities and temperatures of each individual cell change during the expansion.
The chemical equilibrium between the baryon density and quark- and anti-quark densities breaks down at this point. 
Consequently, the chemical  potentials for the single quarks are related to the single quark (and anti-quark) densities 
only and are not connected to the net baryon density. Although we need the chemical potentials for the calculation of the 
energy density, pressure and entropy, these must be calculated from the given quark and anti-quark densities in each cell, 
by using Eqs.~(\ref{chemical_potential_5}) and (\ref{chemical_potential_10}), which determine $\mu_q^i$ and 
${\overline \mu}_q^i$. The net baryon number $n_B$ is conserved and is determined by Eq.~(\ref{baryonic_density_1}) using 
Eqs.~(\ref{chemical_potential_5}) and (\ref{chemical_potential_10}). Only in case of chemical equilibrium 
Eq.~(\ref{baryonic_density_2}) is used to determine the chemical potential $\mu_B$. That is, the $N_B/N_q$ and 
$N_B/{\overline N}_q$ ratio freezes out first at the initialization of our model calculation for the final expansion stage, 
where quarks start to gain mass, the background field $B$ starts to disappear, but the constituent quark numbers do not 
change. The flow does not evolve, the cells expand while coasting with negligible pressure. Still considerable changes 
happen regarding the masses and temperatures of the system.

\section{Non-equilibrium expansion stage}\label{C}

\subsection{Rapid hadronization hypothesis}

A first order phase
transition in chemical and thermal equilibrium with homogeneous 
nucleation would take a long time\cite{CK92}, longer than two particle
correlation measurements indicate. If the homogeneous nucleation
cannot support the required rapid transition then the transition 
becomes delayed and freeze out and hadronization will happen rapidly and 
simultaneously from a supercooled QGP. Thus, a rapid process must
be out of equilibrium, at least of chemical equilibrium \cite{Csorgo_Csernai}.
Possible detailed mechanisms of this out of equilibrium transition are
addressed recently in several works 
\cite{Csernai_Mishustin,Randrup_2010}. 
In the framework of the present model this transition is represented by the 
mass gain and coalescence of constituent quarks, and the simultaneous 
disappearance of the Bag-field. 
These two processes are treated phenomenologically, while we enforce 
conservation laws and our model constraints.

Such a rapid, out of equilibrium process must result in additional 
entropy production \cite{Csorgo_Csernai} from the latent heat of the 
transition (just like at sudden condensation of supercooled vapour). 
We start with our model at the line where we have quarks with 
current quark masses, and the process ends when we reach the 
empirical hadron freeze out line, where quarks have constituent quark mass.

It was found by Cleymans et al. \cite{Cleymans} that, in a wide range of beam energies, detected hadrons freeze out and reach the detectors when the average energy of hadrons (estimated in a thermal, statistical equilibrium fireball model) is between $E_H / N_H =  \left(1.0 - 1.1\right){\rm GeV}$.
Just at the Freeze Out moment the particles do not interact any longer, so the ideal hadron gas mixture is a good approximation. Furthermore, if we plot the FO points on the temperature--chemical potential $[T, \mu_B]$ plane, the
$\left(1.0 - 1.1\right) {\rm GeV}$ constant specific energy contours fall on a continuous line indicating that the final state is representing a statistical thermal equilibrium state of hadrons or a state which is close to it. At the FO state with vanishing interactions and dissipation the expansion of the fireball is adiabatic to a good approximation.

We can also approximate the final FO hadron state by assuming a J\"uttner distribution to all hadrons. 
That means, we use a relation between chemical potential $\mu_B$ and temperature $T$ by means of average energy per hadron 
\begin{eqnarray}
\frac{E_H}{N_H} &=& \frac{\sum\limits_{h=1}^N e_h (m_h,T,\mu_h)}{\sum\limits_{h=1}^N n_h (m_h,T,\mu_h)} = \left(1.0 - 1.1\right) {\rm GeV}\,,
\label{average_energy}
\end{eqnarray}

\noindent
where hadronic energy density and hadronic particle density are given by 
\begin{eqnarray}
e_h &=&\frac{g_h}{4\,\pi^2}\,T \cosh \left(\frac{\mu_h}{T}\right) m_h^3 \left(K_1 \left(\frac{m_h}{T}\right) 
+ K_3 \left(\frac{m_h}{T}\right) \right),
\label{e_h}
\nonumber\\
n_h &=&\frac{g_h}{\pi^2}\,T \sinh \left(\frac{\mu_h}{T}\right) m_h^2 \, K_2 \left(\frac{m_h}{T}\right).
\label{n_h}
\end{eqnarray}

\begin{figure}[!h]
\begin{center}
\includegraphics[scale=0.25]{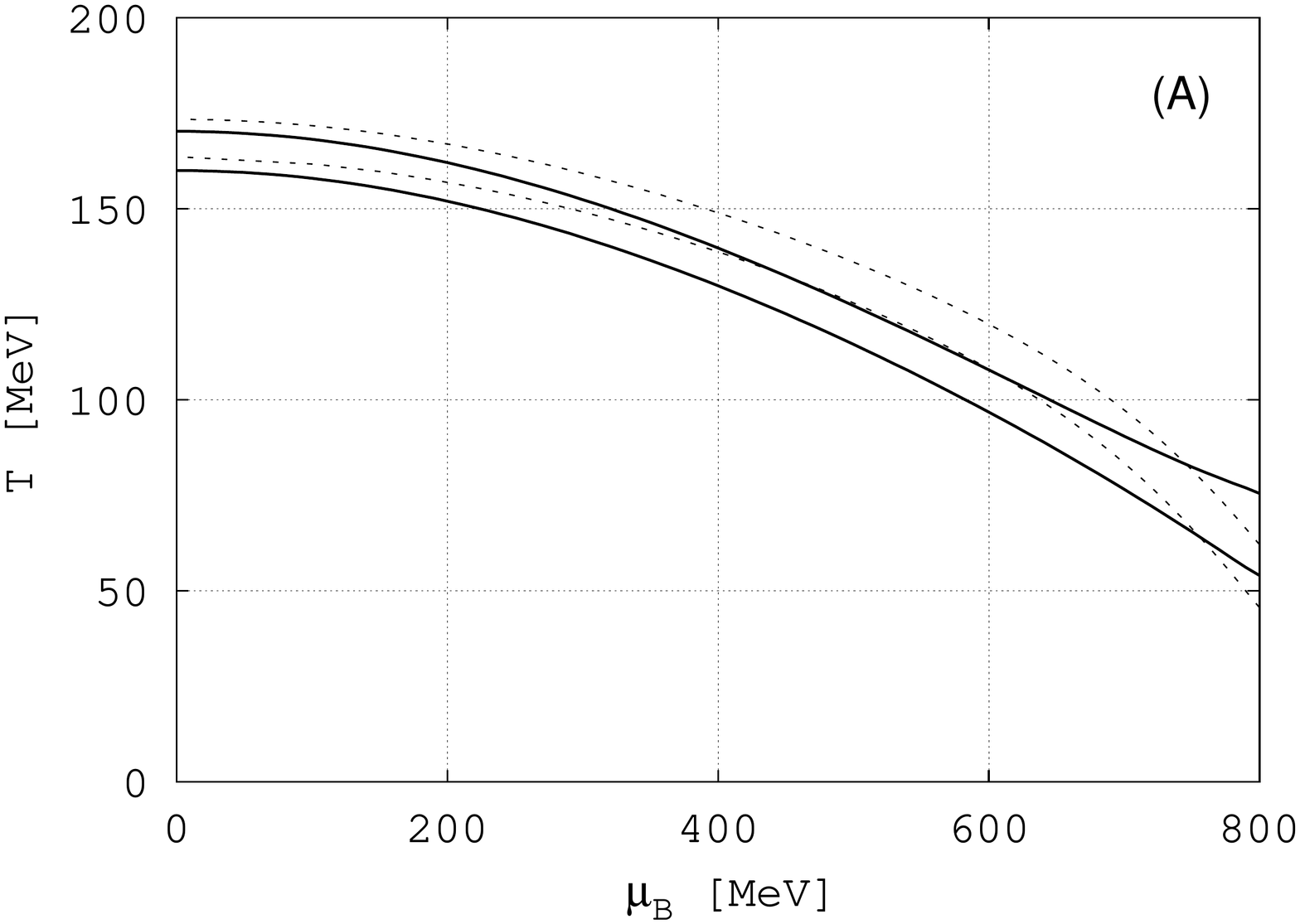}
\includegraphics[scale=0.25]{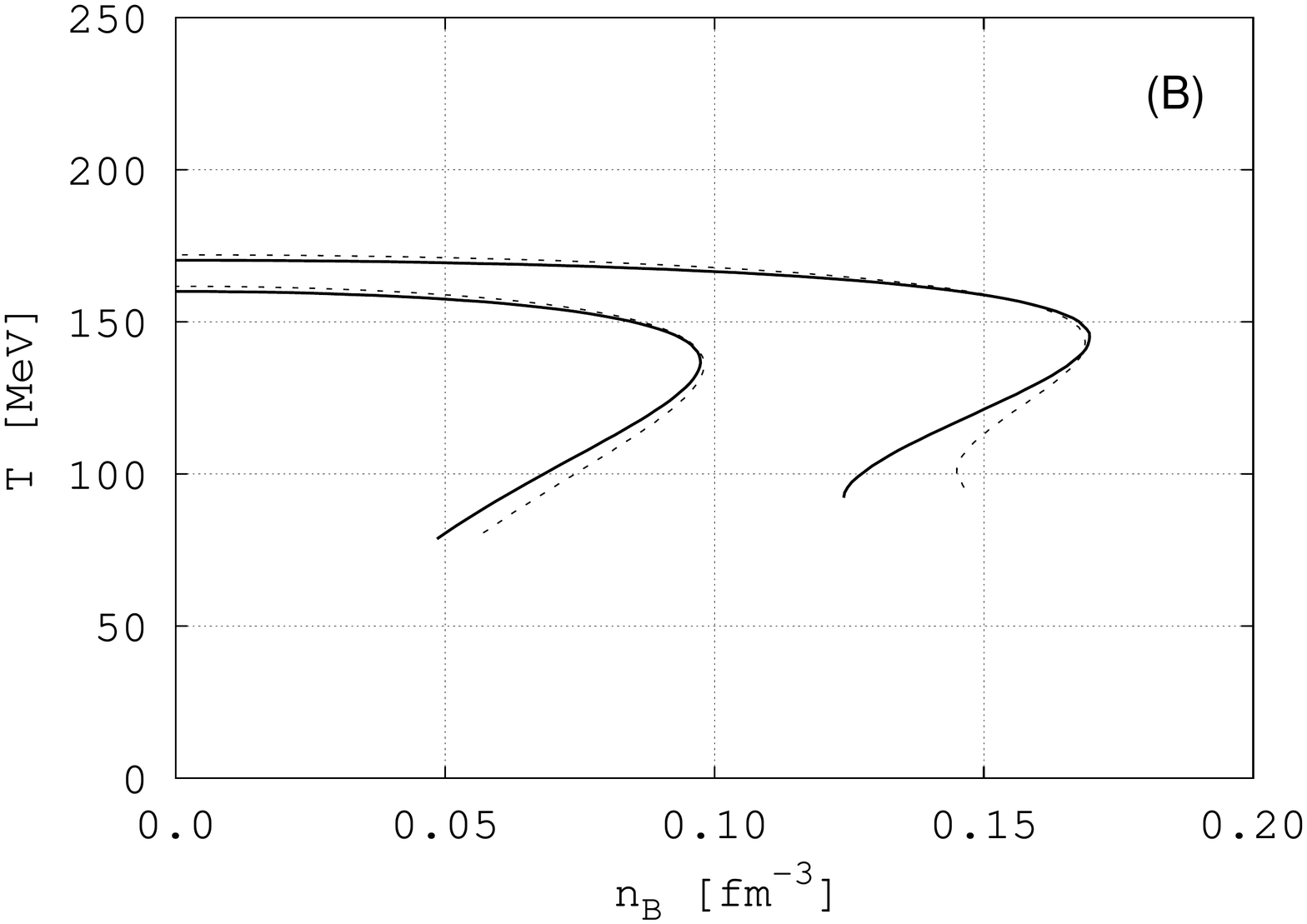}
\end{center}
\caption{Diagram (A) shows the freeze-out curves on the $[\mu_B, T]$-plane, corresponding to mean energy 
$E_{\rm H}/N_{\rm H} = 1.0\,{\rm GeV}$ and $1.1\,{\rm GeV}$ per hadron. The solid lines were computed 
assuming that the hadrons have a J\"uttner distribution, see Eq.~(\ref{average_energy}) (upper solid curve for 
$1.1\,{\rm GeV}$, lower solid curve for $1.0\,{\rm GeV}$).
The dotted lines are from the statistical thermal model \cite{Cleymans} 
(upper dotted curve for $1.1\,{\rm GeV}$, lower dotted curve for $1.0\,{\rm GeV}$).
Diagram (B) shows the same curves transformed to the $[n_B,T]$ plane.}
\label{fig:cleymans-lines}
\end{figure}

\noindent
Here, $m_h$ is the mass of the hadron, $g_h$ is the degeneracy factor, and for baryons we have $\mu_h=\mu_B$ while for mesons $\mu_h=0$.
The Boltzmann-J\"uttner gas approach is a good approximation to the full Fluid Dynamics and Boltzmann equation statistical 
result: in the medium $[T, \mu_B]$-range the FO contour lines are shown in Diagram (A) of FIG.~\ref{fig:cleymans-lines};  
we have taken into account the lightest $100$ hadrons, i.e. $\left(h=\pi^0, \pi^{\pm},K^0,K^{\pm}, ... , D(1950)\right)$. 

In this transition the expansion of the quark gas continues until the average energy per hadron reaches about
$\left(1.0 - 1.1\right){\rm GeV}$.
In a statistical thermal equilibrium model this can be represented on the $[T,\mu_B]$ plane. In the present non-equilibrium model the thermal equilibrium is kept, but $\mu_B$ is not representative of the hadron multiplicity.  However, we can approximate the Cleymans-line by estimating the hadron multiplicity or density after coalescence of our constituent quarks into baryons and mesons.  The freeze-out curve on the $[n_B, T]$ plane is plotted in Diagram (B) of FIG.~\ref{fig:cleymans-lines}.

The cold uniform nuclear matter has an equilibrium density of 
$n_B = \left(1.45 - 1.7\right)$ fm$^{-3}$, similar to the central density of
large nuclei. At higher temperatures a considerable number of mesons are 
also present in statistical equilibrium, thus the {\it effective} baryon
density for the same energy density decreases with increasing temperature.
This enables us to construct an effective Cleymans--freeze-out line on the
$[T,n_B]$-plane for the purposes of our non-equilibrium model.

\begin{figure}[!h]
\begin{center}
\includegraphics[scale=0.25]{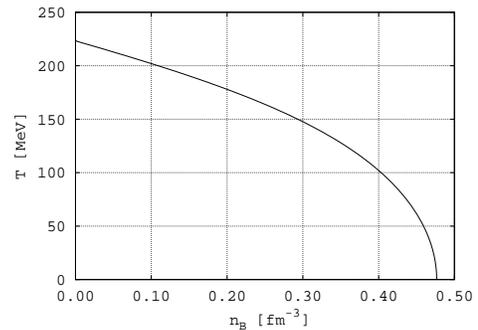}
\caption{The initial conditions for non-equilibrium expansion are determined by vanishing quark condensate, see 
Eqs.~(\ref{Eq_50}) and (\ref{Eq_60}). This corresponds to temperature and density conditions where 
constituent quark mass equals current quark mass $M_f = m_f$ according to Eq.~(\ref{Eq_70}).}
\label{Initial_Condition}
\end{center}
\end{figure}

We have assumed that our rapid hadronization starts at the $[T, n_B]$-contour
where our quarks have current quark mass of about $7\,{\rm MeV}$. This contour is plotted in FIG.~\ref{Initial_Condition}. 
The chemical equilibrium then ceases among quarks and anti-quarks, and we neglect quark and anti-quark annihilations in the 
final expansion stage. Thus, the total number of quarks and anti-quarks, $N_q$ and ${\overline N}_q$, remains constant 
(in contrast to the statistical thermal equilibrium model), while the baryon charge, 
$N_B= \left(N_q - {\overline N}_q\right)/3$, remains also constant as required by baryon conservation.

At the final point of the non-equilibrium expansion quarks will 
coalesce into mesons, baryons and anti-baryons, based on the phase 
space arguments used in \cite{Molnar}. 
The rate of recombination is given by
\begin{eqnarray}
q + \overline{q} \rightarrow m &:& \dot{n}_m  = C_m \frac{g_m}{g_q g_{\bar{q}}} n_q \,{\overline n}_q \,, 
\label{r1} 
\\
q + q + q   \rightarrow b &:& \dot{n}_b  = C_b \frac{g_b}{g_q \,g_q \,g_q} n_q \,n_q \,n_q \,,
\label{r2} 
\\
\overline{q} + \overline{q} + \overline{q}  \rightarrow \overline{b} &:& \dot{\overline n}_b 
= C_b \frac{g_b}{g_q\, g_q\, g_q} {\overline n}_q \,{\overline n}_q \,{\overline n}_q\,.
\label{r3}
\end{eqnarray}

\noindent
Here, $q$ ($\overline{q}$) denotes quarks (anti-quarks), $m$ denotes mesons, $b$ ($\overline{b}$) denotes baryons 
(anti-baryons), $g_m$ and $g_b$ are mesonic and baryonic degeneracy factors, 
and the coefficients $C_m$ and $C_b$ should be 
determined by including phase space factors and an additional normalization constant to satisfy the following relations. 
Notice that the same statistical phase-space factors appear in coalescence rates and in the statistical thermal model 
if the overlap integrals of the coalescing constituents are similar. Eventually other channels via di-quark formation 
can also be taken into account, and the time integrated solution will provide the final hadron abundances.

On the other hand the baryon charge is the excess number of baryons over anti-baryons, and this leads to the normalization requirements for the final  hadrons: 
\begin{eqnarray}
n_B &=& n_b - {\overline n}_b\,.
\label{NN}
\end{eqnarray}

\noindent
Thus, a fraction "$a$" of the anti-quarks may form anti-baryons:
\begin{equation}
{\overline n}_b = a\,\frac{{\overline n}_q}{3}\,,  
\qquad n_b = \frac{n_q - {\overline n}_q}{3} + a\,\frac{{\overline n}_q}{3}\,,
\label{eq:baryon-number}
\end{equation}

\noindent
and the rest of quarks form mesons
\begin{equation}
n_m = \left(1-a\right) {\overline n}_q\,.
\label{eq:meson-number}
\end{equation}

\noindent
The only coefficient, $a$, should arise from the coalescence factors above. 
Initially from Eqs.~(\ref{r2}) - (\ref{r3}) the ratio of formed baryons and 
anti-baryons is $n_b/{\overline n}_b = Q^3$, where $Q \equiv n_q/{\overline n}_q$. 
Assuming that the initial recombination is dominant the conservation laws then yield
\begin{equation}
a \approx \frac{Q-1}{Q^3 - 1}\,.
 \label{eq:recomb-a-param}
\end{equation}

\noindent
The full integration of the rate equations may lead to a change of $a$, which
would modify the ratio of anti-baryons to mesons to a smaller extent.
In \cite{Molnar} the same coalescence model explained the constituent
quark number scaling of the flow parameter, $v_2$, from a weak elliptic
asymmetry of the coalescing quark distributions.

In this way at any stage of expansion, from an initial volume $V_0$ to a point at time $t$ with a given volume $V(t)$, 
we can get all the meson and baryon densities and masses, as well as the total energy from energy conservation 
(neglecting the mechanical work done by the negligible pressure).

From the baryon and meson densities $n_B$ and $n_m$ and the hadronic energy density, $E_H$, we can calculate the baryon and meson chemical potentials (the meson density may exceed the thermal equilibrium value), $\mu_B$ and $\mu_m$, 
and the respective temperatures $T_b(t)$ and $T_m(t)$. Notice that with 
the broken chemical equilibrium the equality of chemical potentials is broken. In this rapid hadronization process the thermal, pressure and flow equilibrium will also break down. Thus the details how this happens should be described in terms of extensive variables as we will see in the next Section.

\subsection{Choice of initial conditions}

Let us consider now the initial conditions for all these parameters.
The applicability of Eq.~(\ref{Eq_70}) is restricted by the condition that the constituent quark mass has to be positive.
Accordingly, we will take the initial conditions such that in the initial state the in-medium chiral condensate $\langle {\overline q} q \rangle_{n_B,T}$ vanishes 
and quarks have a current quark mass. This condition implies, that the density and temperature dependent terms in Eq.~(\ref{Eq_70}) are equal to $1$.
The corresponding curve in the temperature-density plane is shown in FIG.~\ref{Initial_Condition}. 
We used for the current quark mass $m_f= 7 \,{\rm MeV}$, ignoring the difference between $u$ and $d$ quarks.

In this model it is assumed that the chemical equilibrium between the quarks and anti-quarks breaks precisely on this 
curve, and the quark and anti-quark numbers are conserved separately during further expansion. 
At the earlier stage, the quark and anti-quark numbers, $N_{q}$ and ${\overline N}_q$, are computed  from the temperature 
$T$ and baryon number density $n_B$ using Eqs.~(\ref{baryonic_density_2}), 
(\ref{chemical_potential_5}) and (\ref{chemical_potential_10}) with chemical potentials $\mu_q = \mu_B / 3$ and 
${\overline \mu}_q = -\mu_B/3$. But after crossing the line on FIG.~\ref{Initial_Condition} the numbers of quarks and 
anti-quarks, $N_q$ and ${\overline N}_q$, are assumed to stay constant. During further expansion $\mu_q$ and 
${\overline \mu}_q$ change separately and can be obtained numerically from the quark and anti-quark densities, $n_q$ and 
${\overline n}_q$, by means of Eqs.~(\ref{chemical_potential_5}) and (\ref{chemical_potential_10}). 
Since this is a non-equilibrium model, all numerical calculations are carried out on the $[n_B, T]$ plane, using the 
densities $n_B$, $n_q$, ${\overline n}_q$ and the temperature $T$ (and not on the $[\mu_B, T]$ plane).

\subsection{Expansion of the quark gas}

The expansion of the gas of quarks and anti-quarks is considered to be adiabatic, i.e. at constant entropy. 
The initial total entropy density of the gas, $s_0$, can be calculated from the initial temperature, $T_0$, and 
initial baryon charge density, $n_{B}^0$, assuming chemical equilibrium between the quarks and anti-quarks, 
i.e. $\mu_q = - {\overline \mu}_q = \mu_B/3$, and using Eq.~\ref{baryonic_density_2} to get the chemical potential 
and substituting it into Eqs.~(\ref{entropy_10}), (\ref{entropy_11}), and (\ref{entropy_12}).

After the initial moment, the quark and anti-quark densities decrease inversely proportionally to the volume $V$ 
of the system: $n_q = V_0 / V \,n_q^0$, ${\overline n}_q = V_0 / V \, {\overline n}_q^0$, $n_B = V_0 / V \, n_{B}^0$, 
where $V_0$ is the initial volume. Then the total entropy of the system can be expressed simply as a function of
volume, temperature and numerically obtained chemical potentials (or temperature and baryon density)
using Eqs.~(\ref{entropy_10}), (\ref{entropy_11}) and (\ref{entropy_12}).

\begin{figure}[!h]
\begin{center}
\includegraphics[scale=0.25]{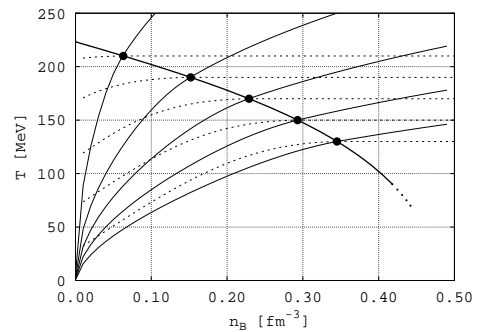}
\end{center}
\caption{Expansion of the gas of quarks and anti-quarks. 
The thick solid line is calculated assuming the quark masses to be equal to the current quark mass, 
that means it is identical with the curve in FIG.~\ref{Initial_Condition}. Beyond that curve we take $M_f = m_f$. 
The dots on the thick solid line indicate the initial temperature and initial baryonic 
density for each curve of adiabatic and iso-ergic expansion. 
The thin solid lines represent the trajectories of the adiabatic expansion of the gas of quarks and anti-quarks. 
The trajectories of the iso-ergic dissipative expansion (dotted lines) are plotted for comparison.}
\label{fig:adiabatic-trajectories}
\end{figure}

Using the condition that the total entropy is constant, i.e. the entropy density also decreases as 
$s(T,V) = s_q(T,V) + {\overline s}_q (T,V) = V_0/V \,s_0$, the expansion trajectories on the $[n_B, T]$ plane can be 
calculated numerically. These trajectories are plotted in FIG.~\ref{fig:adiabatic-trajectories}.

The trajectories corresponding to constant energy expansion (iso-ergic) were calculated, in a similar way, using 
Eqs.~(\ref{Eq_80}) and (\ref{Eq_85}).  This case corresponds to a dissipative expansion. It is apparent from 
FIG.~\ref{fig:adiabatic-trajectories} that the adiabatic expansion where $s=s_0\,n_B/n_B^0$ leads to the fastest 
temperature decrease than the iso-ergic one where $e=e_0\,n_B/n_B^0$.

\subsection{Recombination into hadrons}

\begin{figure}[!h]
\begin{center}
\includegraphics[scale=0.24]{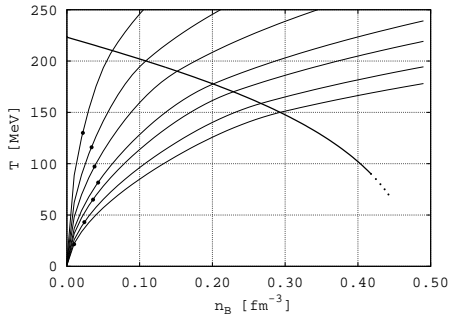}
\end{center}
\caption{A series of curves of adiabatic expansion (thin solid lines) of the gas of quarks and anti-quarks. 
The dots on the adiabatic expansion curves indicate where the rapid freeze-out and hadronization happens. 
These points are determined based on the condition that the energy of the system, including the background field, 
divided by the estimated number of hadrons reaches $1.2 \, {\rm GeV}$ hadron. 
The thick solid line is the same as in FIG.~\ref{Initial_Condition}.}
\label{fig:final-points}
\end{figure}

Due to confining forces, the quarks will finally recombine into hadrons.  We assume that this happens rapidly at the point of recombination when the average energy per hadron (including the background field) decreases to $E_H/N_H = 1.2\,{\rm GeV}$, a value which is still above the 
values of $\left(1.0-1.1\right)\,{\rm GeV}$ obtained by Cleymans et al. \cite{Cleymans}. This corresponds nearly to the energy per hadron of the empirically observed 
freeze-out.  The endpoints of the expansion curves where the recombination happens are shown in FIG.~\ref{fig:final-points}.

The quarks are assumed to recombine into three types of particles: baryons, 
anti-baryons and mesons, which contain three quarks, three anti-quarks, or 
a quark and an anti-quark, respectively.  All hadrons are 
assumed to have a mass that is the sum of the masses of their constituent
quarks at the freeze-out, i.e.
\begin{eqnarray}
M_b &=& {\overline M}_b = 3 \,M_f \left(n_B^{\rm FO}, T^{\rm FO}\right)\,,
\\
M_m &=& 2 \,M_f \left(n_B^{\rm FO}, T^{\rm FO}\right)\,.
\label{eq:hadron-mass}
\end{eqnarray}

\noindent
Further differences between the various hadron species are disregarded.
Most of the anti-quarks will pair with quarks to form mesons, but a small fraction, $a$, will form anti-baryons.  This ratio, $a$, can be estimated based on the recombination rates given in \cite{Molnar}. Thus the baryon, anti-baryon and meson 
densities $n_b$, ${\overline n}_b$ and $n_m$ are calculated from the quark and anti-quark densities 
($n_q$ and ${\overline n}_q$) using Eqs.~(\ref{eq:baryon-number}), (\ref{eq:meson-number}) and (\ref{eq:recomb-a-param}).

\begin{figure}[!h]
\begin{center}
\includegraphics[scale=0.24]{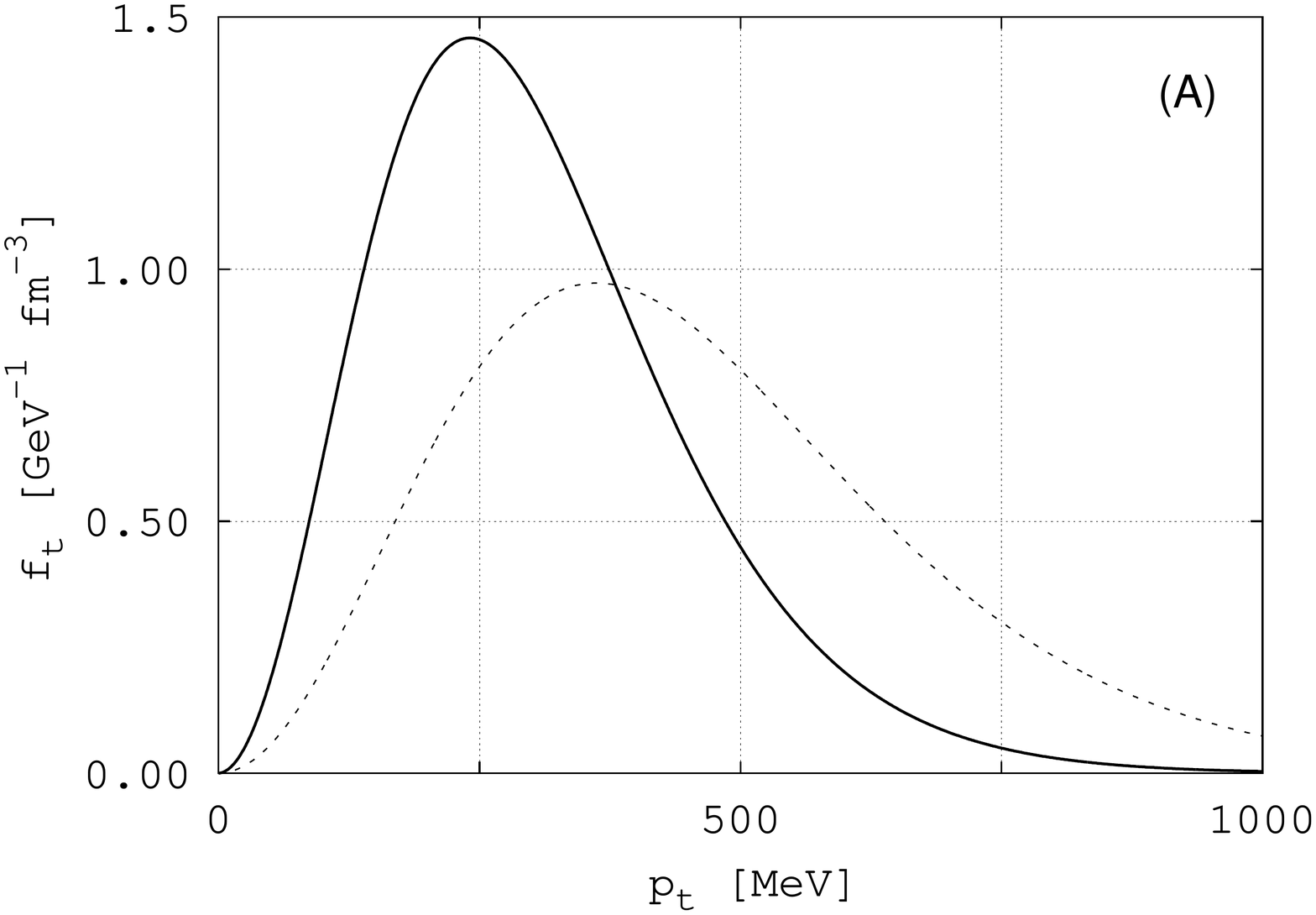}
\includegraphics[scale=0.24]{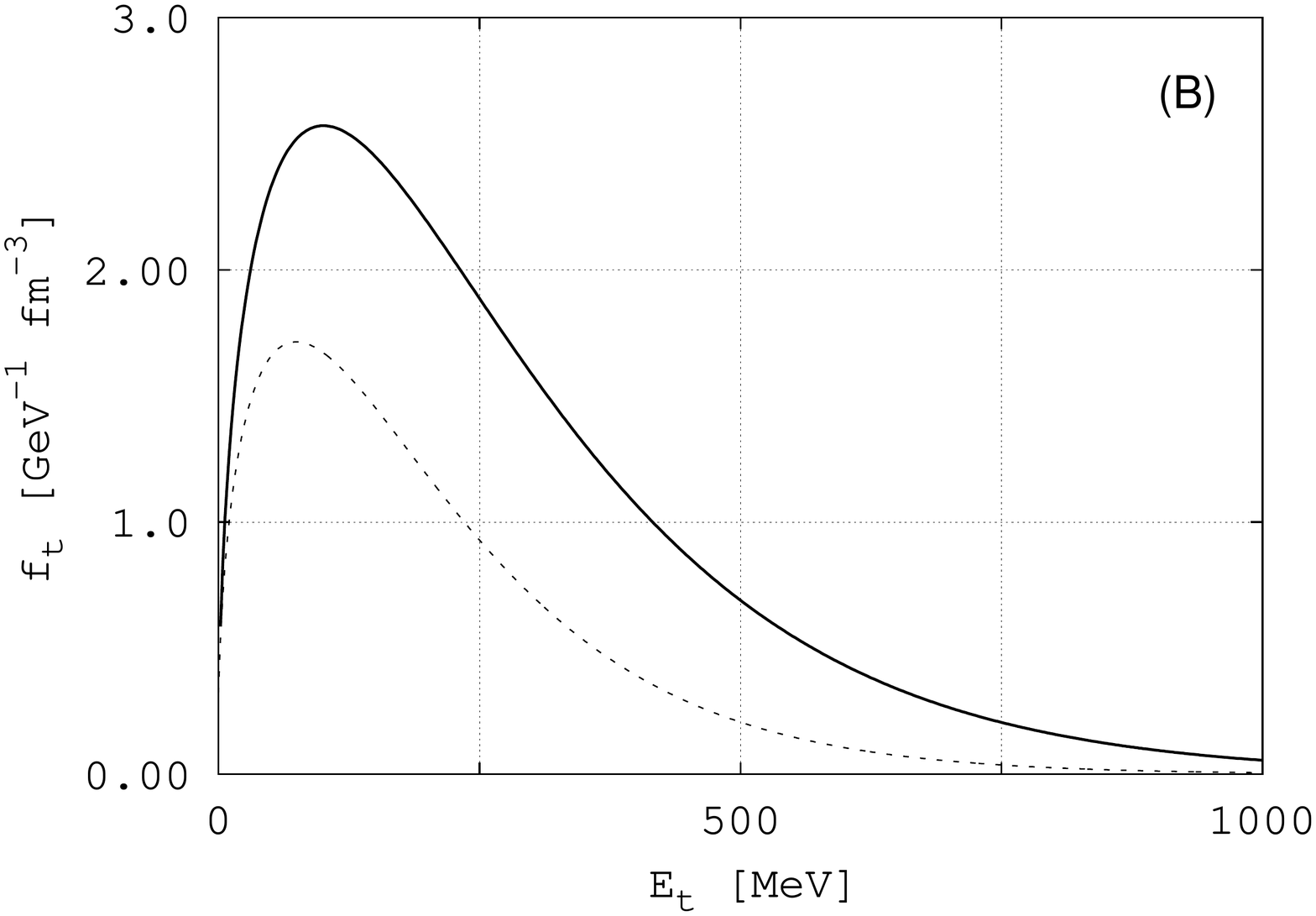}
\end{center}
\caption{Baryon (dotted) and meson (solid) distributions as functions of transverse momentum $p_{\rm t}$ (Diagram (A)) 
and transverse energy $E_{\rm t}$ (Diagram (B)), for the case of an adiabatic expansion. 
The initial state for this calculation was $n_B^0 = 0.21\,{\rm fm}^{-3}$ and $T_0 = 176\,{\rm MeV}$.}
\label{fig:pt-distribution_A}
\end{figure}

\begin{figure}[!h]
\begin{center}
\includegraphics[scale=0.24]{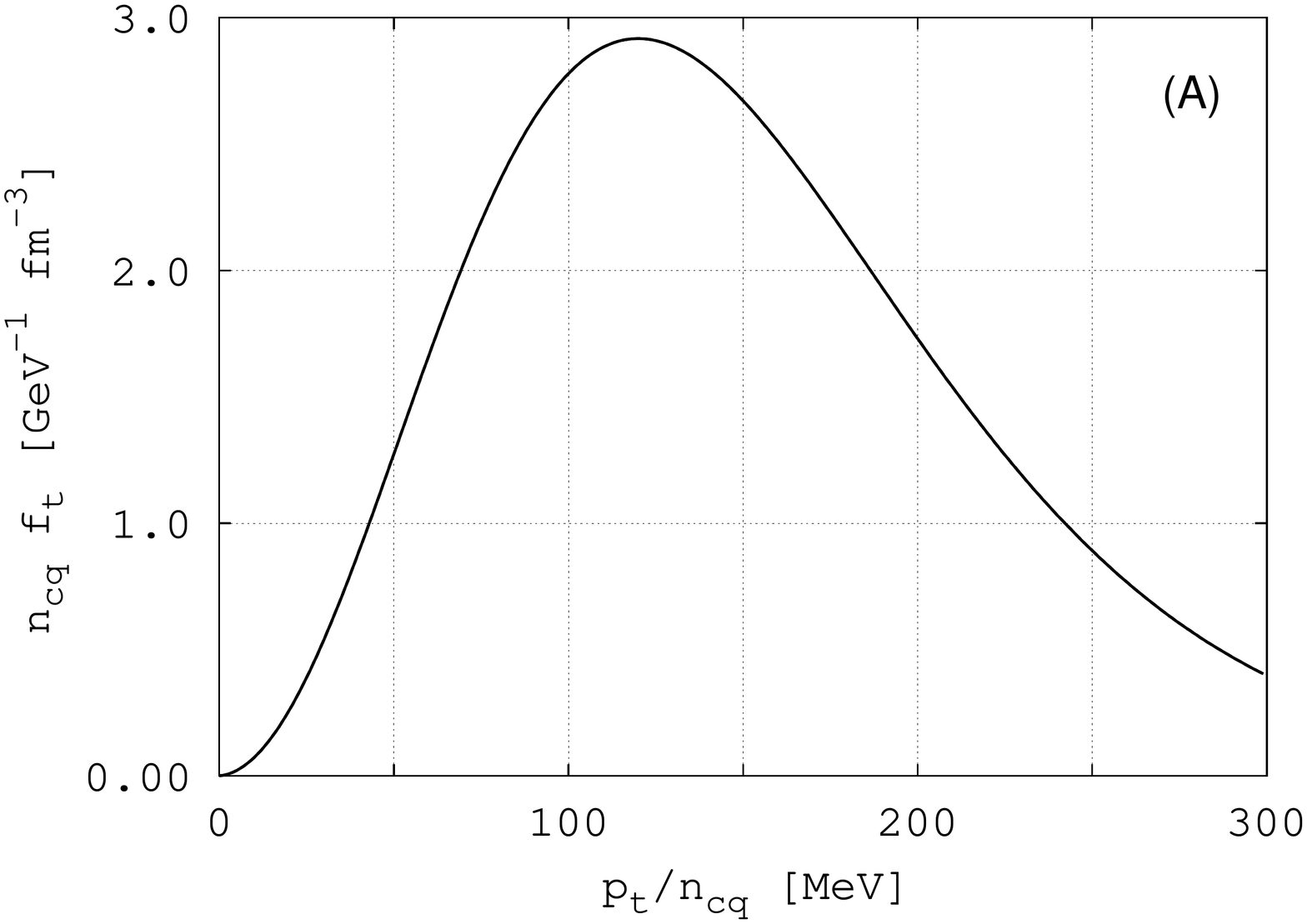}
\includegraphics[scale=0.24]{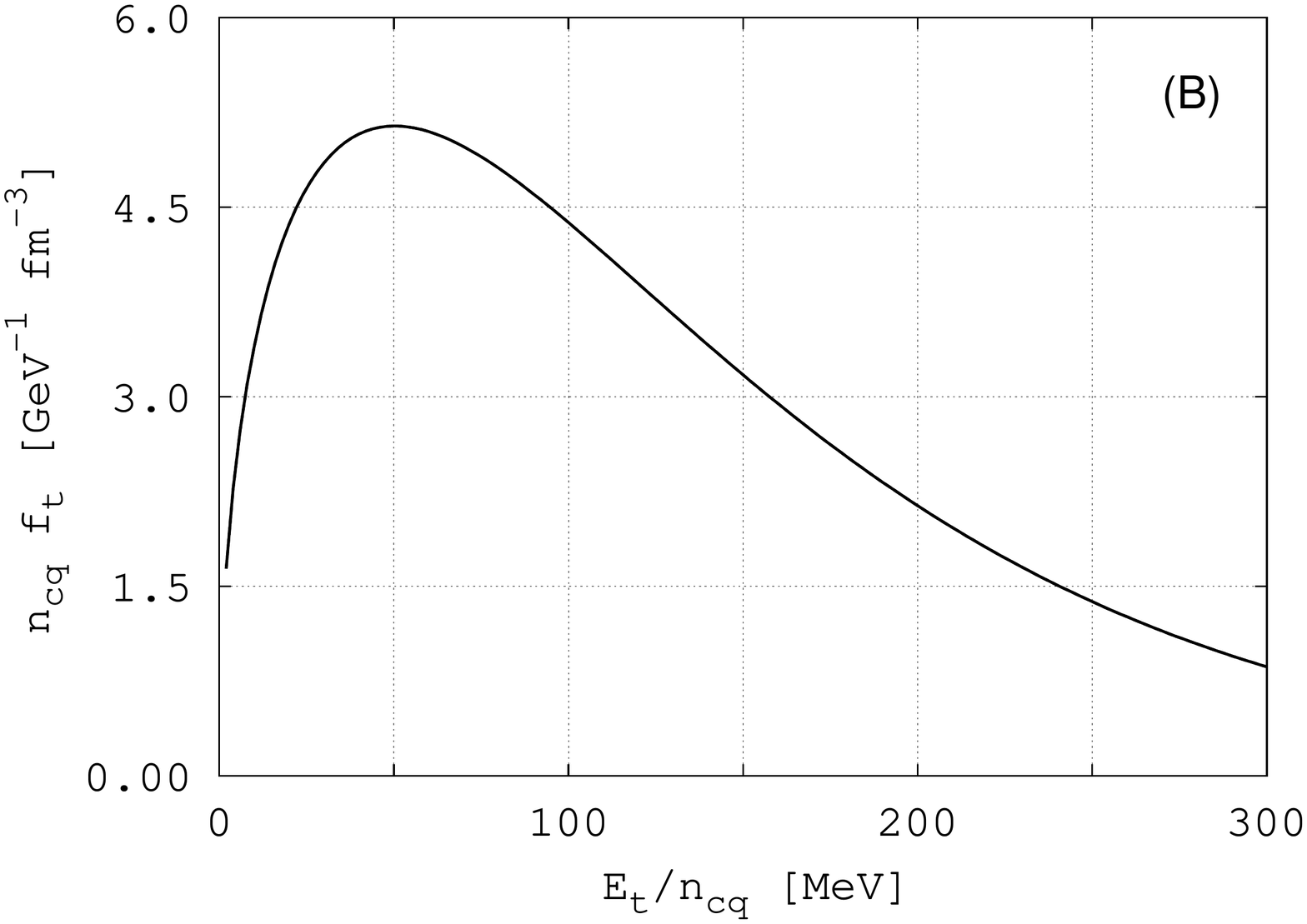}
\end{center}
\caption{These diagrams show the same distribution functions for baryon (dotted) and meson (solid) 
as calculated in FIG.~\ref{fig:pt-distribution_A}, but re-scaled according to constituent quark number $n_{cq}$ 
($n_{cq}=3$ for baryons, $n_{cq}=2$ for mesons). Diagram (A) shows the scaled distribution $n_{cq}\,f_{\rm t}$ 
as function of re-scaled transverse momentum $n_{cq}\,p_{\rm t}$. Diagram (B) shows the re-scaled distribution 
$n_{cq}\,f_t$ as function of re-scaled transverse energy $n_{cq}\,E_{\rm t}$. 
These re-scaled distributions almost coincide for baryons and mesons; they are on top of each other so no 
difference can be seen here in the diagrams.  
The initial state for this calculation was $n_B^0 = 0.21\,{\rm fm}^{-3}$ and $T_0 = 176\,{\rm MeV}$.}
\label{fig:pt-distribution_B}
\end{figure}

We assume that at the point of recombination the flow freezes out too, and 
both the thermal and chemical equilibria cease. 
The hadrons are assumed to have J\"uttner distribution after the freeze out, 
but the temperature parameter in this distribution will be different for 
baryons and mesons. The parameters of the distribution after recombination
are calculated from the condition of energy conservation: the thermal energy
of each hadron type will be equal to the energy of their constituent quarks,
$E^{\rm th}_m = E^{\rm th}_b$.
Due to the different masses of baryons and mesons, their temperature parameters
will be different. The temperature ratio $T_b / T_m$ will correspond to the
mass ratio $M_b / M_m = 3/2$.  The distributions of baryons and mesons for a
calculation done with initial state $n_B^0 = 0.21\,{\rm fm}^{-3}$
and $T_0 = 176\,{\rm MeV}$ is shown in FIG.~\ref{fig:pt-distribution_A} 
and FIG.~\ref{fig:pt-distribution_B}.
Here the final baryon and meson temperatures are $T_b = 228\,{\rm MeV}$ and
$T_m = 152\,{\rm MeV}$ and $T_b / T_m = 3/2$, while the final constituent quark
mass is 308 MeV. The scaled $p_{\rm t}$ and $E_{\rm t}$ distributions become
identical under this condition, however this is not enough to reproduce
the NCQ scaling of $v_2(p_{\rm t})$ indicating that the recombination
influences the flow velocities of the final hadrons. This concept was
already pointed out in \cite{Molnar} based on the properties of the
collision integral. It is important to point out that the transport
theoretical treatment and the collision integral are applicable also
at situations when the local equilibrium has ceased to exist.

\section{Elliptic flow}\label{D}

The elliptic flow parameter, $v_2$, can be calculated from the final, post Freeze-Out distribution by the 
Cooper-Frye formula. Assuming an isochronous FO hypersurface, we obtain simple expressions for final 
measurables \cite{Elliptic_Flow1}.

\subsection{Formula for the elliptic flow}

In this Section we mainly follow the arguments of Refs.~\cite{Elliptic_Flow1,Elliptic_Flow2}.
The kinematic average of a quantity $A (\ve{x},\ve{p})$ is given by
\begin{eqnarray}
\langle A \rangle &=& \frac{\displaystyle \int d^3 x\;\int d^3 p\; f(\ve{x},\ve{p})\; 
A (\ve{x},\ve{p})}{\displaystyle \int d^3 x\;\int d^3 p\;f(\ve{x}, \ve{p})}\,,
\label{Eq_1}
\end{eqnarray}

\noindent
where $f (\ve{x}, \ve{p})$ is the one-particle distribution function.  Especially, we are interested in the elliptic flow $v_2$, that is the
following kinematic average
\begin{eqnarray}
v_2 &=& \left\langle \frac{p_x^2 - p_y^2}{p_{\rm t}^2}\right\rangle .
\label{Eq_5}
\end{eqnarray}

\noindent
According to Eqs.~(\ref{Eq_1}) and (\ref{Eq_5}), we obtain
\begin{eqnarray}
v_2 &=&
\frac{\displaystyle \int d^3 x\;\int d^3 p\; f(\ve{x},\ve{p})\; \frac{\displaystyle p_x^2 - p_y^2}{\displaystyle p_{\rm t}^2}}
{\displaystyle \int d^3 x\;\int d^3 p\;f(\ve{x}, \ve{p})}\;.
\label{Eq_10}
\end{eqnarray}

\noindent
Now we subdivide the system into $N$ fireballs or cells, each of which has a given volume $V_i$ and contains a given number 
of particles $N_i$, which are distributed by a given distribution function $f_i (\ve{x},\ve{p})$ inside cell $i$. 
Since in a given cell the elliptic flow parameter $v_2$ does not depend on coordinate $\ve{x}$, we can take the integral 
over $d^3 x$ and obtain
\begin{equation}
v_2 =
\frac{\displaystyle 
\sum\limits_{i=1}^{N}\;V_i\;
\int\limits_{0}^{2\pi} d \phi 
\int\limits_0^{\infty} d p_{\parallel} \int\limits_0^{\infty} d p_{\rm t} \;p_{\rm t} \;
f^i (\ve{p})\; \cos 2 \phi}
{\displaystyle 
\sum\limits_{i=1}^N\;V_i\;
\int\limits_{0}^{2\pi} d \phi 
\int\limits_0^{\infty} d p_{\parallel} \int\limits_0^{\infty} d p_{\rm t} \;p_{\rm t} \;
f^i (\ve{p})}\,,
\label{Eq_15}
\end{equation}

\noindent
where we have used the relation
$\cos 2 \,\phi = \cos^2 \phi - \sin^2 \phi = 
\frac{\displaystyle p_x^2}{\displaystyle p_{\rm t}^2} - 
\frac{\displaystyle p_y^2}{\displaystyle p_{\rm t}^2}$ 
with 
$p_{\rm t} = \sqrt{p_x^2 + p_y^2}$ 
being the transverse momentum. 
In our study we consider the mid-rapidity particles, 
($p_z = 0$ or the rapidity $y = 0$), 
i.e.\ $\ve{p}= (p_x,p_y)$ and there is no integration 
for the longitudinal momentum or rapidity.
Furthermore, we are interested in the momentum dependence of 
elliptic flow parameter $v_2$. Thus we obtain from (\ref{Eq_15}) 
the expression:
\begin{equation}
v_2 \left(p_{\rm t}, y=0\right)  =
\frac{\displaystyle 
\sum\limits_{i=1}^{N}\;V_i\;
\int\limits_{0}^{2\pi} d \phi\; f^i(p_x,p_y)\; \cos 2 \phi}
{\displaystyle 
\sum\limits_{i=1}^N\;V_i\;
\int\limits_{0}^{2\pi} d \phi\;f^i(p_x,p_y)}\,,
\label{Eq_16}
\end{equation}

\noindent
where the sum runs over all individual cells $i=1,...,N$, and we have used the fact that for mid-rapidity particles $\ve{p}^2 = p_x^2 + p_y^2 = \ve{p}_{\rm t}^2$.  Let us study the elliptic flow of a single type of particle. We assume a J\"uttner distribution in each individual cell of the bag, see Eq.~(\ref{Juttner_A}), and obtain
\begin{eqnarray}
f_{\rm J}^i &=&
\frac{1}{\left(2\,\pi\,\hbar\right)^3}\,{\rm exp}\left(\frac{\mu^i-\gamma_i\,p_0^i+\gamma^i\,v_x^i\,p_x 
+ \gamma^i\,v_y^i\,p_y}{T_i}\right) 
\nonumber\\
&=& \frac{1}{\left(2\,\pi\,\hbar\right)^3}\,{\rm exp} \left(\frac{\mu^i-\gamma_i\,p_0^i+\gamma^i\,v^i\,p_v }{T_i}\right),
\label{Juttner}
\end{eqnarray}

\noindent
where $p_v$ is the component of the momentum parallel to the cell velocity $\mathbf{v}^i$, and $\mu^i$ is the chemical potential of the given particle type in the cell $i$; here we have also used $u^{\mu}_i = \gamma^i (1, v_x^i, v_y^i, 0)\,,\quad u_{\mu}^i = \gamma^i (1, - v_x^i, - v_y^i, 0)$ and 
$p^{\mu} = (p_0, p_x, p_y, 0)\,, \quad p_{\mu} = (p_0, - p_x, - p_y, 0)$. Since the elliptic flow parameter is calculated after chemical freeze-out has 
happened, each particle type has their own chemical potential
\begin{equation}
\frac{1}{(2\pi \hbar)^3} \,{\rm exp} \left({\frac{\mu^i}{T}}\right) = \frac{n_i}{4\pi\,M_i^2\,T_i\,K_2(M_i/T_i)}\,,
\label{eq:juttner-density}
\end{equation}

\noindent
where $n_i$ is the density and $M_i$ is the mass of the given particle type. 
For particles at CM rapidity the zeroth component of four-momentum equals the transverse mass, 
i.e. it is just the energy of one particle
\begin{eqnarray}
p_0^i = \sqrt{M_i^2 + p_{\rm t}^2} = M^i_{\rm t}\,.
\label{transverse_mass}
\end{eqnarray}

\noindent
Then, according to Eqs.~(\ref{Eq_16}) and (\ref{Juttner}), we obtain the following expression for the elliptic flow:
\begin{eqnarray}
v_2 &=& \frac{\displaystyle \sum\limits_{i=1}^{N}\;V_i\;\int\limits_0^{2\,\pi}\; d \phi\;\cos 2 \phi\;f^i_{\rm J} (p_x,p_y)}
{\displaystyle \sum\limits_{i=1}^{N}\;V_i\;\int\limits_0^{2\,\pi}\; d \phi\;f^i_{\rm J} (p_x,p_y)}\,,
\label{elliptic_flow_A}
\end{eqnarray}

\noindent
where the sum runs over all cells of the bag. Now let us insert Eq.~(\ref{Juttner}) into expression 
(\ref{elliptic_flow_A}). Then, by means of Eq.~(\ref{eq:juttner-density}) we obtain the following expression for the 
$v_2$ parameter:
\begin{equation}
\displaystyle
v_2 =\frac{\displaystyle \sum_{i=1}^N \tilde N_i\,
e^{-\gamma^i M^i_{\rm t} / T_i} \int_0^{2\pi} d\phi \; \cos 2\phi \; e^{\gamma^i \,v_i\, p_v / T_i}}
{\displaystyle \sum_{i=1}^N \tilde N_i\,e^{-\gamma^i \,M^i_{\rm t} / T_i} 
\int_0^{2\pi} d\phi \; e^{\gamma^i \,v_i \,p_v / T_i}}\,,
\end{equation}

\noindent
where $\tilde N_i$ denotes
\begin{equation}
  \tilde N_i = V_i\,\frac{n_i}{T_i\,K_2(M_i/T_i)}.
  \label{eq:n-tilde}
\end{equation}

\noindent
If the direction of the velocity of cell $i$ relative to axis $x$ is denoted by $\phi_0^i$, then $p_v$ can be written as
$p_v = p_{\rm t} \cos (\phi - \phi_0)$, and the following expression is obtained:
\begin{equation}
v_2(\pt) = \frac{\displaystyle \sum_{i=1}^N \tilde N_i\,
e^{-\gamma^i M_{\rm t}^i / T_i} \,\cos 2\phi^i_0 \; I_2(\gamma^i \,v_i\, p_{\rm t} / T_i)}
{\displaystyle \sum_{i=1}^N \tilde N_i\,e^{-\gamma^i M_{\rm t}^i / T_i} \; I_0(\gamma^i \,v_i\, p_{\rm t} / T_i)}\,.
\label{eq:v2-full}
\end{equation}

\noindent
$I_0$ and $I_2$ denote the zeroth and second order modified Bessel functions of the first kind \cite{Abramowitz_Stegun}.

Notice that if the source temperatures are all equal or if the Freeze Out
temperature of all fluid cells is the same, than one observes a
linear increase of $v_2$ as a function of $\pt$ or $\pt/ n_{cq}$. This
is frequently quoted as a linear hydrodynamical increase, observed already
in the first model calculations, e.g. \cite{Elliptic_Flow1}.  Such linear increase
leads trivially to a constituent quark number scaling. It is easy to see
that this feature is just a consequence of the equal temperature assumption.
If for example we assume a large, hotter, static central source, the high $\pt$
behaviour of $v_2$ will be dominated by this hot source and thus
$v_2$ will decrease  at high $\pt$ as pointed out in ref. \cite{Elliptic_Flow2}.

If we assume that the temperatures of all cells are the same, $T_i = T$, then the temperature- and mass-dependent parts 
of ${\tilde N}_i$ (see Eq.~(\ref{eq:n-tilde})) cancel from the numerator and denominator of Eq.~(\ref{eq:v2-full}), 
and we get (note at FO the constituent quark mass does not depend on density and temperature anymore)
\begin{equation}
v_2(\pt) =
\frac{
\displaystyle \sum_{i=1}^N N_i e^{-\gamma^i M_{\rm t}^i / T} \; \cos 2\phi^i_0 \; I_2(\gamma^i \,v_i \,p_{\rm t} / T)}
{\displaystyle \sum_{i=1}^N N_i e^{-\gamma^i M_{\rm t}^i / T} \; I_0(\gamma^i \,v_i \,p_{\rm t} / T)}\,.
\label{eq:v2-simple}
\end{equation}

\noindent
In the special case of four cells moving into the four directions ($\pm x, \pm y$), Eq.~(\ref{eq:v2-simple}) 
reduces to the expression given in \cite{Elliptic_Flow1}.

The simplest configuration that can approximate elliptic flow is dividing the system into two cells that move in opposite directions with the same velocity $v$.  For this ``two-cell" model case the $v_2$ parameter is expressed as
\begin{equation}
  v_2(\pt) = \frac{I_2(\gamma v \pt / T)}{I_0(\gamma\, v \, \pt / T)}\,.
  \label{eq:v2-twocell}
\end{equation}

\noindent
A slightly more complicated possibility is having one larger non-moving central cell, and two smaller side cells moving in opposite directions, as shown in FIG.~\ref{fig:three-cells}.  In this ``three-cell" model case $v_2(\pt)$ can be expressed as
\begin{equation}
v_2(\pt) = \frac{2 N_s\, e^{-\gamma M_{\rm t}/T} I_2(\gamma\, v \, \pt / T)}{2 N_s\, e^{-\gamma M_{\rm t}/T} I_0(\gamma \, v \, \pt / T) 
+ N_c\, e^{-M_{\rm t}/T}}\,.
\label{eq:v2-threecell}
\end{equation}

\noindent
Here $N_c$ denotes the particle number of the central cell, while $N_s$ denotes the particle number of the identical side-cells.  This configuration corresponds to a flow with less pronounced asymmetry.  The large central cell has a spherical momentum distribution, while the smaller side cells introduce a slight asymmetry to this.

\subsection{Calculation of the elliptic flow}

\subsubsection{Two-cell model}

\begin{figure}[!h]
\begin{center}
\includegraphics[scale=0.25]{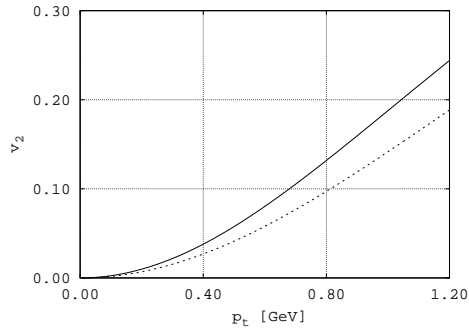}
\end{center}
\caption{The $v_2$ parameter as a function of \pt, calculated from the two-cell model, Eq.~(\ref{eq:v2-twocell}). 
The initial state used is the same as in FIG.~\ref{fig:pt-distribution_A} and FIG.~\ref{fig:pt-distribution_B}. 
The dotted curve represents the baryons, while the solid curve represents the mesons. 
The cell velocities for baryons and mesons are $v_b = 0.26$ and $v_m = 0.21$, corresponding to a flow-energy ratio of 
$3/2$ of the constituent quarks of the two different particle types (calculated relativistically).}
\label{fig:v2-pt}
\end{figure}

As a first approximation, the $v_2$ parameter was calculated using a simple model of elliptic flow where the system is divided into two droplets moving in opposite directions with the same velocity.  For this model, $v_2$ is given by
Eq.~(\ref{eq:v2-twocell}). The baryons and mesons were given different flow energies, such that the ratio of flow energy 
per quark is $(FE_b / n_{cq}) / (FE_m / n_{cq}) = 3/2$.
This leads to different flow velocities for baryons and mesons, and reproduces the constituent quark number scaling of the
elliptic flow parameter. In this model the scaling according to the transverse momentum \pt and transverse energy
$E_{\rm t}$ are tied to each other, i.e. the scaling is either present or not for both these
variables. By this point of the evolution of our system, all energy in the background field is exhausted and the internal, excitation and random kinetic energies of the hadrons reach the FO value $\left(1.0 - 1.1\right)\,{\rm GeV}$.
The $v_2(\pt)$ curve obtained with the initial state parameters discussed in the previous Section is shown in FIG.~\ref{fig:v2-pt}. 

\begin{figure}[!h]
\begin{center}
\includegraphics[scale=0.25]{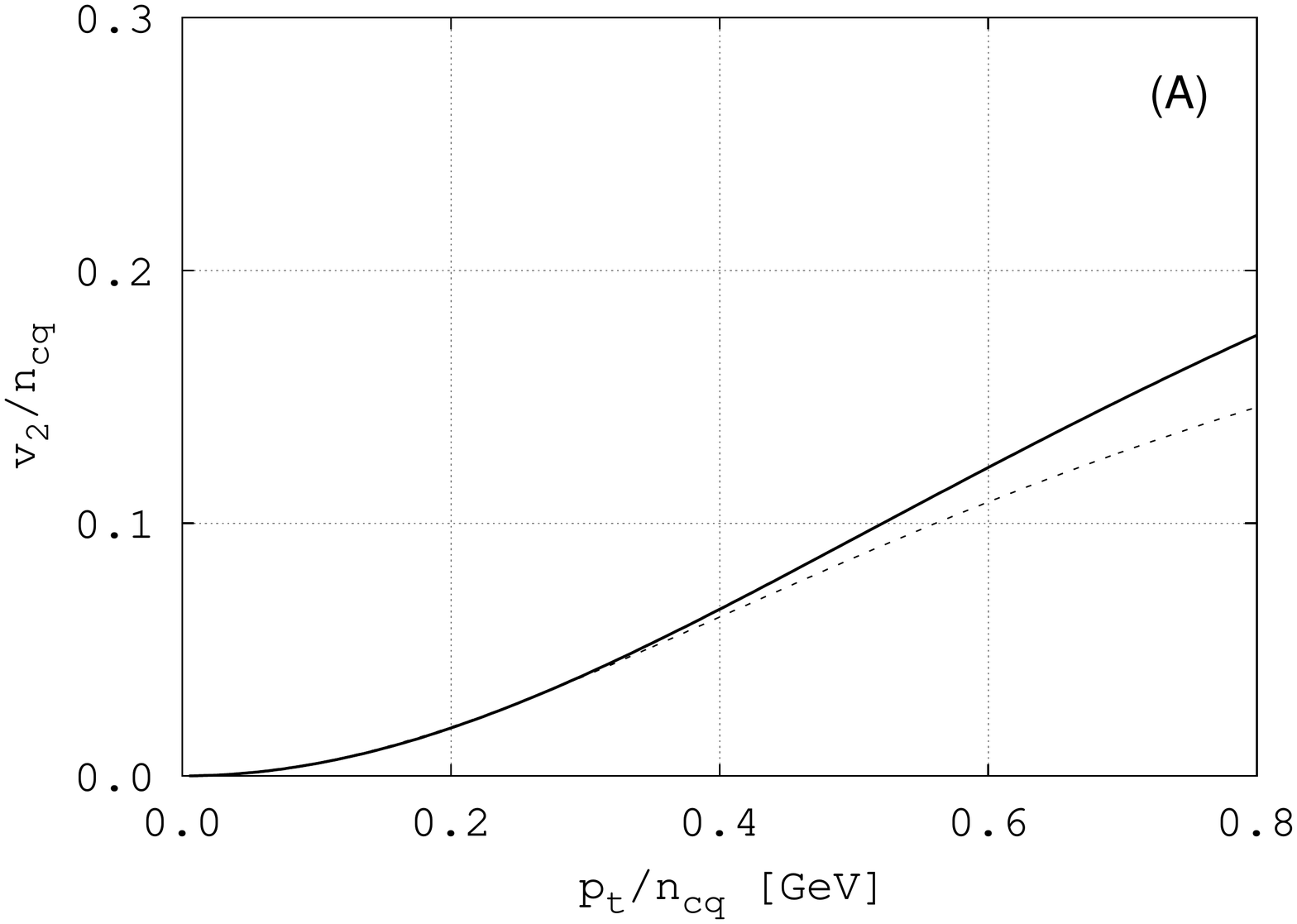}
\includegraphics[scale=0.25]{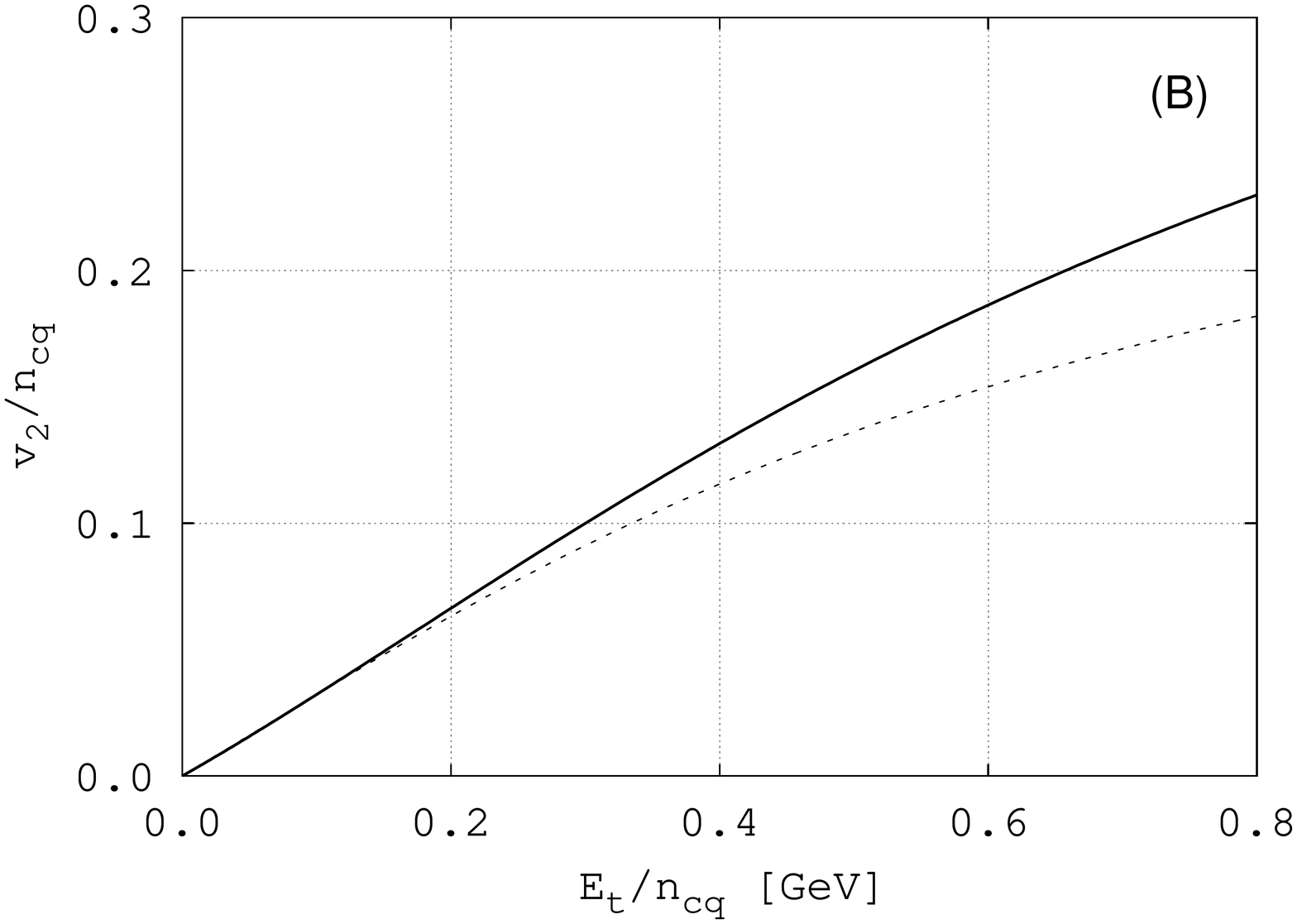}
\end{center}
\caption{The re-scaled elliptic flow parameter, $v_2 / n_{cq}$, as a function of $\pt / n_{cq}$ (Diagram (A)) and 
$E_{\rm t} / n_{cq}$ (Diagram (B)), calculated from the two-cell model. The dotted curves represent the baryons while the 
solid curves represent the mesons.  The curves coincide for low \pt\ value, i.e. the constituent quark number scaling 
is reproduced for the low \pt\ region.}
\label{fig:v2-cqn}
\end{figure}

The elliptic flow parameter, re-scaled according to constituent quark number is shown in FIG.~\ref{fig:v2-cqn}.  In this figure, the baryon- and meson-curves coincide for low \pt, i.e.\ constituent quark number scaling of $v_2$ is reproduced for small values of the transverse momentum, but not for $\pt > 400$ MeV.
These results of QNS can be further improved by considering a three-cell model, as we will see in the next Subsection. 

\subsubsection{Three-cell model}

\begin{figure}[!h]
\begin{center}
\includegraphics[scale=0.25]{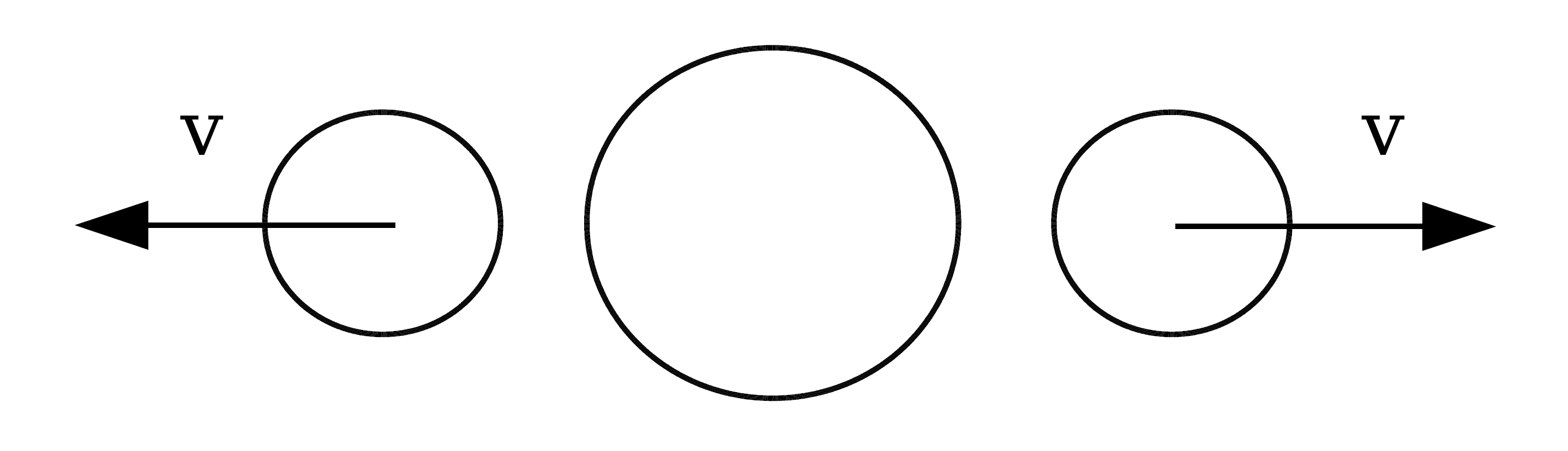}
\end{center}
\caption{Scheme of three-cell model.
The asymmetric flow is approximated by dividing the system into either two cells that move in opposite
directions with velocity $v$, or two smaller moving side cells and a stationary central cell.
The schematic of the latter configuration is shown here.}
\label{fig:three-cells}
\end{figure}

The asymmetry in the two-cell model described in the previous Subsection is very strong. In \cite{Molnar} it is shown that
the constituent quark number scaling is more precise if the $v_2$ coefficient is small, and the higher harmonic
coefficients, $v_k$ ($k>2$) are negligible. Therefore, a three-cell model with one large stationary
central cell and two moving side cells was also studied. The schematic scheme of this arrangement of cells is shown on
FIG.~\ref{fig:three-cells}.

\begin{figure}[!h]
\begin{center}
\includegraphics[scale=0.25]{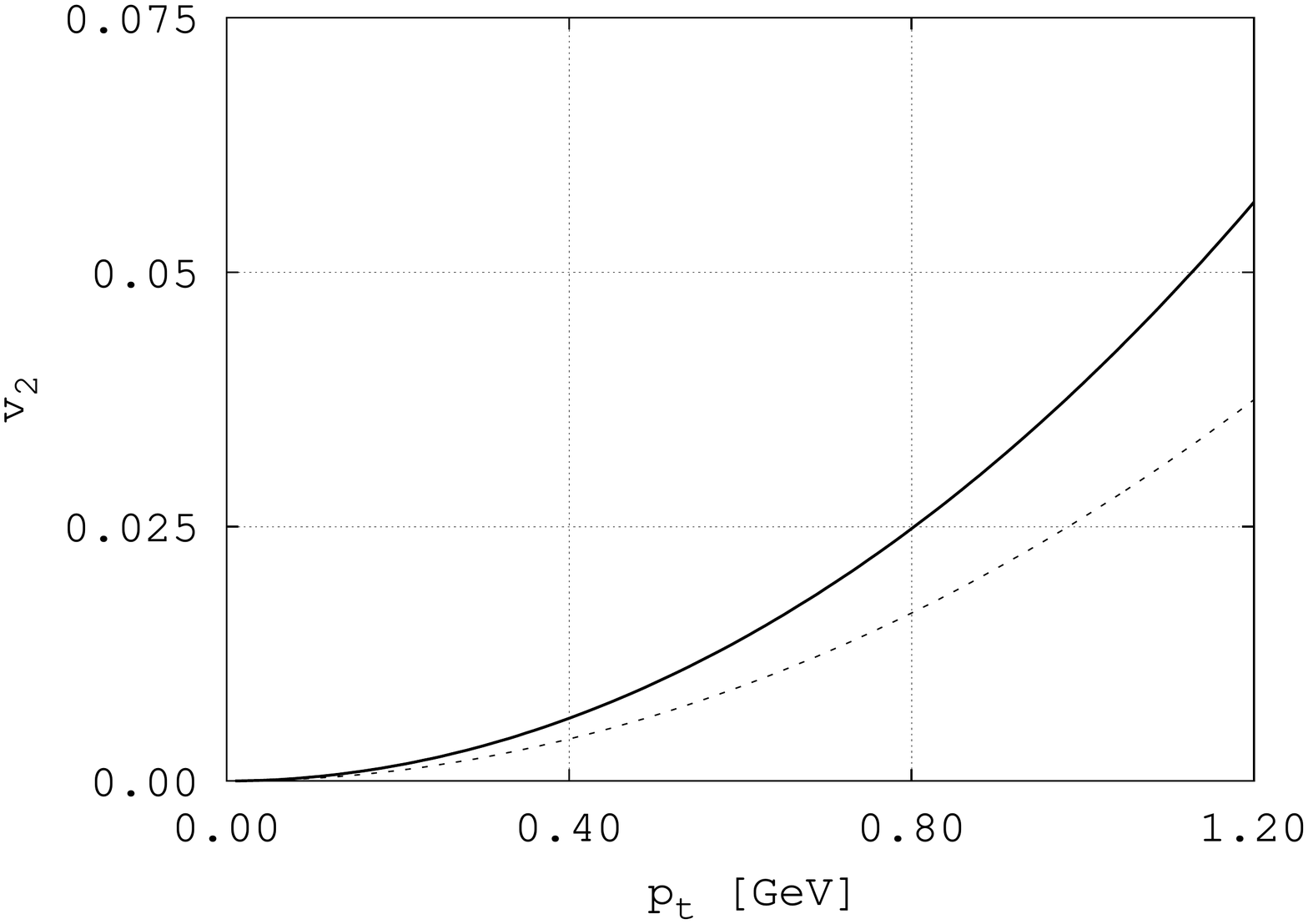}
\end{center}
\caption{The $v_2$ parameter as a function of \pt, calculated from the three-cell model, Eq.~(\ref{eq:v2-threecell}).
The initial state used is the same as in FIG.~\ref{fig:pt-distribution_A} and FIG.~\ref{fig:pt-distribution_B}.
The dotted curve represents the baryons, while the solid curve represents the mesons.
The cell velocities for baryons and mesons are $v_b = 0.26$ and $v_m = 0.21$, corresponding to a flow-energy ratio of
$3/2$ of the constituent quarks of the two different particle types (calculated relativistically). 
A particle ratio $N_c / N_s = 10$ has been assumed.}
\label{fig:v2-pt-three}
\end{figure}

\begin{figure}[!h]
\begin{center}
\includegraphics[scale=0.25]{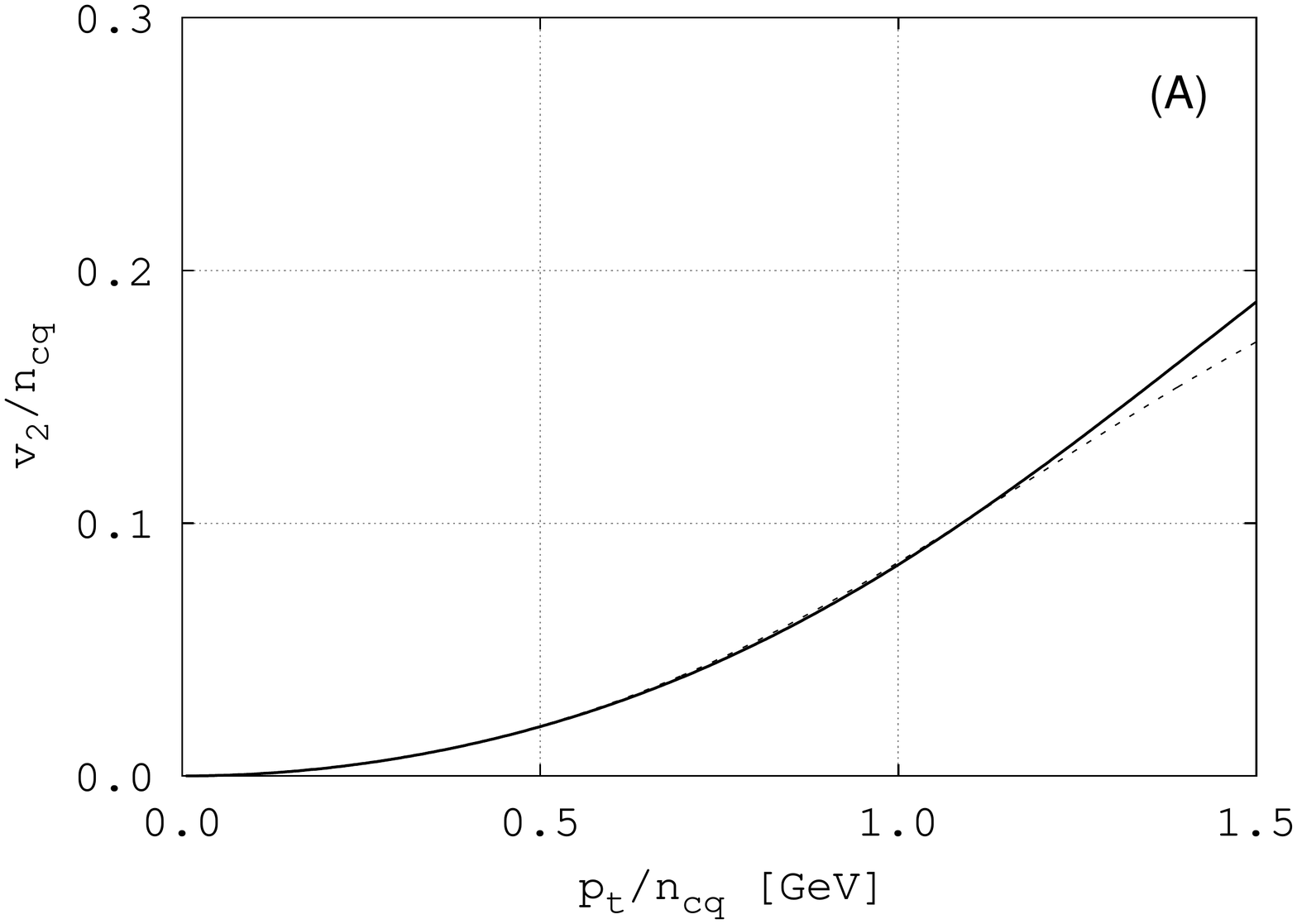}
\includegraphics[scale=0.25]{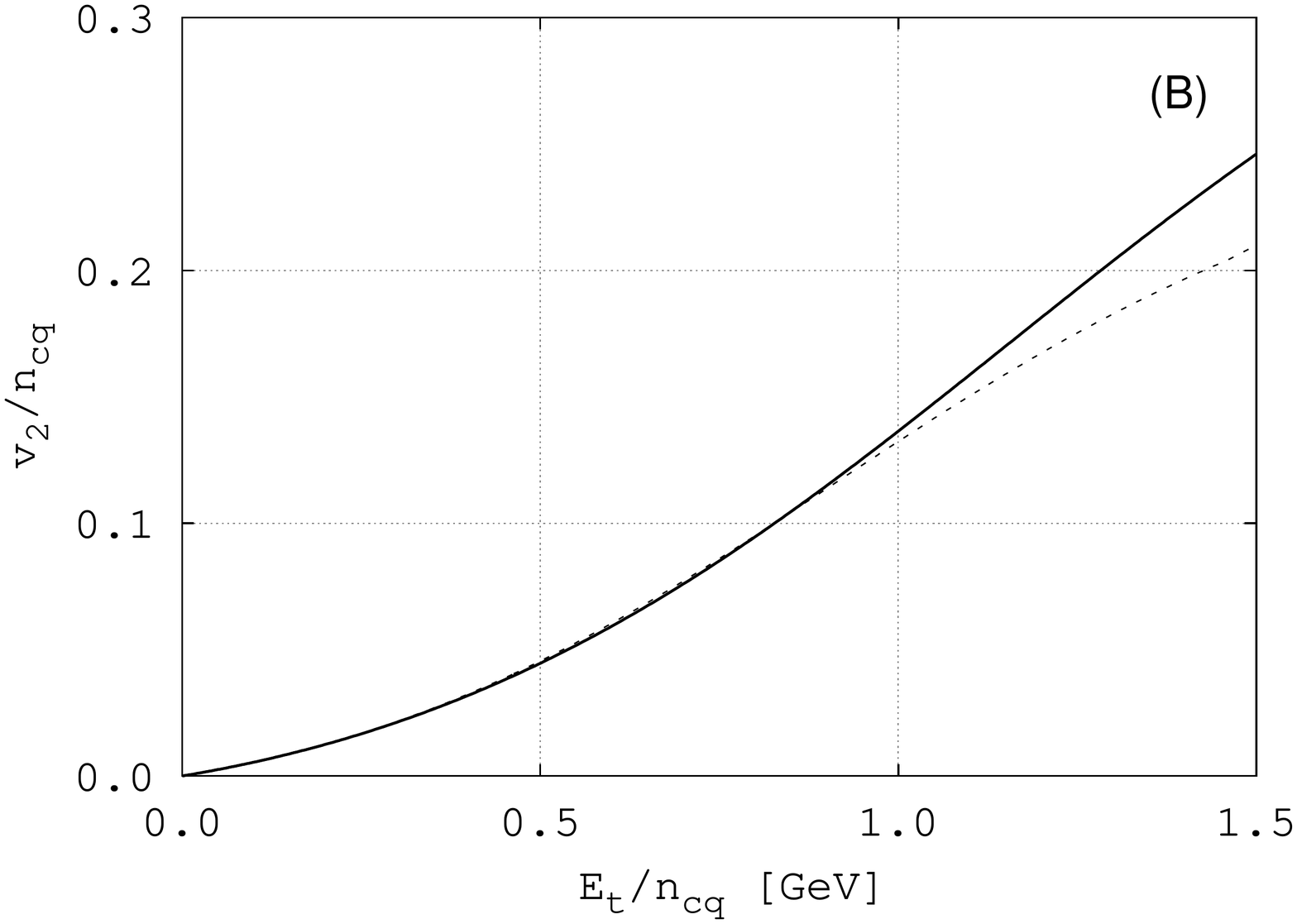}
\end{center}
\caption{The re-scaled elliptic flow parameter, $v_2 / n_{cq}$, as a function of $\pt / n_{cq}$ (Diagram (A)) and
$E_{\rm t} / n_{cq}$ (Diagram (B)), calculated from the three-cell model. The dotted curves represent the baryons while 
the solid curves represent the mesons.  The curves overlap nearly completely, and the 
constituent quark number scaling is reproduced for a much wider range 
of \pt\ than in the case of the two-cell model. The velocities of the side sources 
are the same as in FIG.~\ref{fig:v2-pt}. A particle ratio $N_c / N_s = 10$ has been assumed.}
\label{fig:v2-3-cqn}
\end{figure}

The elliptic flow parameter of the three-cell model is given by Eq.~(\ref{eq:v2-threecell}) 
and the result is shown in FIG.~\ref{fig:v2-pt-three}. 
The side cells were assumed to have the same velocities as in the case of the two-cell model, 
and the temperature of all cells was the same. The particle number ratio of the central cell to the side cells was set 
to $N_c / N_s = 10$. It should be noticed, that the elliptic flow parameter, shown in FIG.~\ref{fig:v2-pt-three}, 
is not insensitive on such choice, e.g. a ratio of $N_c / N_s = 5$ would enlarge the value of $v_2$. 

The elliptic flow parameter, re-scaled according to constituent quark number is shown in FIG.~\ref{fig:v2-3-cqn}. 
We have found that QNS is insensitive on the chosen particle number ratio $N_c / N_s = 10$, 
e.g. a ratio $N_c / N_s = 5$ yields very similar results as shown in FIG.~\ref{fig:v2-3-cqn} 
According to the results in FIG.~\ref{fig:v2-3-cqn}, the three-cell model is able to reproduce the constituent quark number 
scaling of $v_2$ for a wider range of \pt\ values compared to the two-cell model. It must be noted that this is not only 
a result of the reduction of the momentum distribution asymmetry compared to the two-cell model. 
However, the choice of flow velocities for baryons and mesons is still relevant.

\section{Summary}\label{E}

A model of rapid hadronization was considered in which the effective
constituent quark mass depends on density and temperature. In this model the evolution starts 
from an ideal QGP in chemical, thermal and mechanical equilibrium. Then, one after the
other the chemical, thermal and mechanical equilibrium break down rapidly, while the
quarks build up constituent quark mass, and the background gluon field (bag constant) breaks up
and vanishes. This model can be considered as a simple representation of the breaking chiral symmetry
and deconfinement in a dynamical transition crossing the Quarkyonic phase \cite{McLerran_Pisarski}.

The elliptic flow parameter $v_2$ was calculated for the final hadron distributions obtained from the model, and the constituent quark number scaling 
was partially reproduced. We assume the following stages of hadronization: 1) The chemical equilibrium between the quarks and anti-quarks breaks and the chiral symmetry breaks at the same time. 2) The quark and anti-quark numbers are assumed to be conserved during further expansion, i.e. quark, anti-quark pair creations and annihilations are ceased. 3) As the quark gas expands and cools, the quarks gain mass according to Eq.~(\ref{Eq_70}), and the field 
associated with the Bag constant $B$ decreases. 4) The thermal freeze-out and recombination into hadrons are completed when the mean energy per hadron reaches the FO value of $E_H/N_H~=\left(1.0-1.1\right){\rm GeV}$. The created hadrons are not in local thermal and flow equilibrium with each other.  We considered only three types of hadrons: baryons, anti-baryons and mesons, ignoring the differences between different hadron species, and the momentum distribution of hadrons was calculated.

Using the obtained momentum distributions, the elliptic flow parameter $v_2$ was computed from a simple multi-source model 
of the elliptic flow. Two cases were considered. In the first case, the flow asymmetry is approximated by two fireballs 
moving in opposite directions. Although this leads to a very strong asymmetry in the momentum distribution, it leads to a 
$v_2(\pt)$ curve that scales with the constituent quark number for small \pt\ values only.  In the second case, 
we added a large central fireball to reduce the asymmetry and approximate pure elliptic flow better (i.e. reduce the 
higher $v_k$ ($k>2$) harmonic coefficients). This three-source model was able to reproduce the constituent quark number 
scaling for a wide range of \pt\ values.

The present model is highly simplified and attempts to provide an
insight to the rapid hadronization and freeze out process
in view of the quark number scaling. We considered the features arising from
breaking down of equilibrium in this process in terms of thermo and
fluid dynamical parameters which are applicable to partial components
of the matter. We intend to implement these concepts in more complex models (like hybrid models) 
and to search for more fundamental reasons for the observed freeze out features
in terms of the partial extensives.

Despite the simplicity of the model, it is capable of reproducing the constituent quark number scaling of the $v_2$ elliptic flow parameter.  The presence of constituent quark number scaling in experimental data suggests that the elliptic flow develops in the Quark-Gluon Plasma phase, before the quarks recombine into hadrons. Therefore understanding the origin the elliptic flow can provide insight into the quark phase of matter.

\section*{Acknowledgements}\label{G}

Enlightening discussions with Prof. Daniel D. Strottman are gratefully acknowledged. The authors also would like to thank 
Dr. Kalliopi Kanaki, Yun Cheng, Dr. Csaba Anderlik and Dr. Etele Molnar for useful hints and advices. 
Sven Zschocke thanks for the sincere hospitality of the Physics Department of University of Bergen, and Magne H\r{a}v\r{a}g 
is acknowledged for kind computer assistance. This work was supported by the Alexander von Humboldt Foundation, 
by the Meltzer Fund of the University of Bergen and by the Computational Subatomic Physics Project at Uni-Research of the 
Research Council of Norway. Igor N. Mishustin acknowledges partial support from DFG grant 436RUS 113/957/0-1 (Germany) and 
grants NS-7235.2010.2 and RFBR09-02-91331 (Russia).

\end{document}